\documentclass[prd,aps,preprint,nofootinbib,showpacs,superscriptaddress]{revtex4}
\usepackage{graphicx}
\usepackage{hyperref}
\usepackage{amsfonts}
\usepackage{amsmath,amssymb}
\usepackage{bm}
\usepackage{color}
\usepackage{epstopdf}
\usepackage{epsfig}
\usepackage{float}

{\rm }

\def\be{\begin{equation}}
 \def\ee{\end{equation}}
 \def\bea{\begin{eqnarray}}
 \def\eea{\end{eqnarray}}

\def\A{\mathcal{A}}

\def\2{\frac{1}{2}}
\def\4{\frac{1}{4}}

\catcode`\@=11

\def\@normalsize{\@setsize\normalsize{15pt}\xiipt\@xiipt
\abovedisplayskip 14pt plus3pt minus3pt%
\belowdisplayskip \abovedisplayskip
\abovedisplayshortskip  \z@ plus3pt%
\belowdisplayshortskip  7pt plus3.5pt minus0pt}
\def\small{\@setsize\small{13.6pt}\xipt\@xipt
\abovedisplayskip 13pt plus3pt minus3pt%
\belowdisplayskip \abovedisplayskip
\abovedisplayshortskip  \z@ plus3pt%
\belowdisplayshortskip  7pt plus3.5pt minus0pt
\def\@listi{\parsep 4.5pt plus 2pt minus 1pt
            \itemsep \parsep
            \topsep 9pt plus 3pt minus 3pt}}

\def\underline#1{\relax\ifmmode\@@underline#1\else
        $\@@underline{\hbox{#1}}$\relax\fi}
\@twosidetrue \relax
\def\be{\begin{equation}}
 \def\ee{\end{equation}}
 \def\bea{\begin{eqnarray}}
 \def\eea{\end{eqnarray}}
 \def\bes{\begin{eqnarray}}
 \def\ees{\end{eqnarray}}
 \def\bi{\begin{itemize}}
 \def\ei{\end{itemize}} 

\catcode`@=12

\catcode`\@=11

\def\ps@headings{\def\@oddfoot{}\def\@evenfoot{}
\def\@oddhead{\hbox{}\hfill
        \makebox[.5\textwidth]{\raggedright\ignorespaces --\thepage{}--
        \hfill }}
\def\@evenhead{\@oddhead}
\def\subsectionmark##1{\markboth{##1}{}}
}

\ps@headings

\catcode`\@=12

\begin{document}

\title{Holographic Fermi Liquids in a Spontaneously Generated
Lattice }

%

\author{James Alsup}
\email{jalsup@umflint.edu}
\affiliation{Computer Science, Engineering and Physics Department,
The University of
Michigan-Flint,
Flint, MI 48502-1907, USA}
\author{Eleftherios Papantonopoulos}
\email{lpapa@central.ntua.gr}
\affiliation{
 Department of Physics, National Technical University of
Athens,
Zografou Campus GR 157 73, Athens, Greece}
\author{George Siopsis}
\email{siopsis@tennessee.edu}
\author{Kubra Yeter\footnote{Present Address: Department of Chemistry and Physics, Coastal Carolina
University, P.O. Box 261954, Conway, SC 29528-6054, USA }}
\email{kyeter@tennessee.edu}
\affiliation{
 Department of Physics and Astronomy, The
University of Tennessee, Knoxville, TN 37996 - 1200, USA
}

\date{\today}


\begin{abstract}
We discuss  fermions in a spontaneously generated
holographic lattice background.  The lattice structure
at the boundary is generated by introducing a
higher-derivative interaction term between a $U(1)$ gauge field
and a scalar field. We solve
the equations of motion below the critical temperature at which the lattice forms, and analyze the change in the
Fermi surface due to the lattice. The
fermion band structure is found to exhibit a gap due to
lattice effects.
\end{abstract}
\maketitle
\section{Introduction}

The normal state of   copper oxide high-temperature superconductors has been found to possess
anomalous thermal and optical conductivities and electrical resistivity. An
explanation of this behavior was put forward in \cite{Abrahams}, citing that over a wide range of
momenta, there exist excitations contributing to
both the charge and spin polarizability.  The  retarded
one-particle self-energy accounting for the exchange of charge and spin
fluctuations were found to be quite different than that of a conventional Fermi liquid.
The  spectral function $A(k,
\omega )$  is much broader and
carries substantially more weight in the wings because of an
$\omega^{-1}$ tail.
The
behavior is referred to as a
marginal Fermi liquid.

The formation of a marginal Fermi liquid in the normal phase of the
superconductor implies that the quasiparticle excitations are
unstable. A way to stabilize the quasiparticles
 was proposed in \cite{Faulkner:2009am} by employing
the gauge/gravity duality in which strongly-coupled phenomena were
 studied using dual, weakly-coupled gravitational systems
\cite{Maldacena:1997re,Gubser:2002,Witten:1998}.
A coupling  was introduced between the fermion and the
condensate formed in the superconducting phase capable of stabilizing the quasiparticles with a
gap. A similar investigation was carried out in
\cite{Gubser:2009dt}, where the effect of a background, zero-temperature
superconducting anti-de Sitter domain wall was considered.

A system with interacting bosonic and fermionic degrees of freedom
at finite density was considered holographically in
\cite{Nitti:2014fsa}. By computing the two-point Green function of
the boundary fermionic operator, it was shown that in the presence
of the charged scalar condensate, the dual field theory exhibited
electron-like and/or hole-like Fermi surfaces. Compared to
fluid-only solutions, the presence of the scalar condensate
destabilized the Fermi surfaces with lowest Fermi momenta and a
gap appeared.

Recently, in the context of holography, lattice effects on the
Fermi surfaces have garnered interest. The study of condensed matter systems
on the boundary in the presence of a lattice requires a
gravitational background lacking spatial homogeneity.
 Lattice effects were
introduced in a strongly-coupled  system of fermions at a finite
density in \cite{LSSZ:2012}. The  holographic dual consisted  of
fermions in the presence of a
Reissner-Nordstr\"om-anti-de Sitter black hole with the lattice effect encoded by
 periodic modulation of the chemical potential with a wavelength
on the order of the Fermi momentum.  In the marginal Fermi liquid regime a gap was formed
due to the interaction between different excitation levels.  The spectral weight remained small but non-zero inside of the gap. This behavior is described as a ``pseudogap'' to differentiate from that of a band gap with identically zero spectral weight.

A  Dirac field in the presence of a holographic lattice was studied in \cite{Ling:2013fl}. In the low temperature limit, the Fermi surface was modified by
lattice effects from a circle to an ellipse.  The behavior was
attributed to the presence of quasiparticles made more
massive by renormalization effects due to the lattice.  Additionally, a pseudogap band structure was found at
the intersection of the Fermi surface and the Brillouin zone
boundary.

The  holographic lattice background may be constructed by introducing
a scalar field with  periodic boundary conditions along a
spatial direction or by employing  a periodic chemical potential
for the scalar potential of the gauge field. In most of the cases,
a spatial inhomogeneity is introduced perturbatively
\cite{Maeda:2011pk,LSSZ:2012,Aperis:2010cd,
Flauger:2010tv,Hutasoit:2012ib,Erdmenger:2013zaa,Donos:2013eha}.
Recently, the authors in \cite{Horowitz:2012ky} and
\cite{Horowitz:2012gs}  constructed some spatially
inhomogeneous but periodic gravitational backgrounds by fully
solving the coupled partial differential equations numerically
with the Einstein-DeTurck method.
In \cite{Alsup:2013kda}, it was shown that by turning on a higher-derivative
interaction term between a $U(1)$  gauge field and a scalar
field,  the scalar field spontaneously develops a spatially-dependent profile.  Through backreaction, the charge density spontaneously developed
spatial inhomogeneity.

In this work, we introduce a Dirac field in a gravity
background. A spatial inhomogeneity and subsequently a modulated charge density are spontaneously
generated, in the
spirit of \cite{Alsup:2013kda}.  We study the boundary lattice effects on
the Fermi surface by analytically solving, up to second order,
the backreacted geometry and Dirac equation
in the bulk. The
structure of the Fermi surface and a pseudogap 
behavior of the fermions are analyzed with the Green
function behavior of the Dirac field.

The paper is organized as follows. In Section \ref{fermisetup}, we perturbatively
study the numerical and analytic solutions of a holographic
system with
 spontaneously generated inhomogeneous phases introduced by higher
 derivative couplings. In Section \ref{Diracsection},
we introduce a Dirac field into the holographic lattice system. We
analyze the lattice effects on the Fermi surface by
calculating the spectral function of the system.
 To this end, we
perturbatively solve the Dirac equations in the periodic background at small but finite
temperature. We discuss the supporting numerical
results for the spectral function at the critical temperature (zeroth order) in Section \ref{DiracZeroth}.
In Section \ref{DiracFirst} we present the solutions below the critical temperature, the first order in Section \ref{DiracFirstAna}, and  the second order of
perturbation  in Section \ref{DiracSecondAna}. In Section \ref{bandgapAna} we discuss the generation of the pseudogap.
Finally, in Section \ref{DiracConclusion}, we discuss the results
of this study and present an outlook for future studies.


\section{General Formalism}
\label{fermisetup}

We begin by considering a holographic system consisting of a $U(1)$ gauge field $A_\mu$, of
field strength $F_{\mu\nu}=\partial_\mu A_\nu-\partial_\nu A_\mu$,
and of a scalar field $\phi$ of charge $q$ under the $U(1)$ group of
the gauge field, having  mass $m$. Later, a Dirac field with mass $m_f$ and
charge $q_f$ will be added to the system. The $AdS$ spacetime geometry
of the bulk where these fields live has a negative cosmological
constant of $\Lambda=-3/L^2$.

We consider the following action
\begin{equation}
S=\int d^4x\sqrt{-g}\mathcal{L}~, \ \ \ \
\mathcal{L}=\frac{R+6/L^2}{16 \pi
G}-\frac{1}{4}F_{\mu\nu}F^{\mu\nu}-(D_\mu \phi)^*D^\mu \phi-m^2
|\phi|^2~,\label{actionorig}
\end{equation}
with $D_\mu \phi=\partial_\mu\phi-iq A_\mu \phi$. In the rest of
this paper,  we shall choose units so that $16 \pi G = L =1$.

Following \cite{Alsup:2013kda}, we introduce higher derivative
terms that lead to spatial inhomogeneity in the boundary theory. These terms were shown to lead to a holographic lattice structure on the boundary,
\be \mathcal{L}_{\mathrm{int}} = \eta
\mathcal{G}^{\mu\nu} (D_\mu \phi)^\ast D_\nu \phi - \eta' | D_\mu
\mathcal{G}^{\mu\nu} D_\nu \phi |^2 \ , \label{int_Lag_den}\ee
where $\mathcal{G}_{\mu\nu} = F_{\mu\rho}{F_{\nu}}^{\rho}-\frac{1}{2}g_{\mu\nu}F^{\rho\sigma}F_{\rho\sigma}$.

The action \eqref{actionorig} with the additional interaction term
\eqref{int_Lag_den} gives the Einstein equations
\begin{equation}
G_{\mu\nu}  - 3 g_{\mu\nu} = \frac{1}{2} T_{\mu\nu}
~,\label{Eineq}
\end{equation}
where $T_{\mu\nu}$ is the stress-energy tensor,
\begin{equation}
T_{\mu\nu} = T_{\mu\nu}^{(EM)} + T_{\mu\nu}^{(\phi)} + \Theta_{\mu\nu}~,
\end{equation}
and the gauge, scalar, and interaction contributions, respectively, may be written as
\bea T_{\mu\nu}^{(EM)} &=&F_{\mu\rho}{F_{\nu}}^{\rho}-\frac{1}{4}g_{\mu\nu}F^{\rho\sigma}F_{\rho\sigma}~,  \nonumber\\
T_{\mu \nu}^{(\phi)} &=& (D_\mu\phi)^\ast D_\nu \phi +  D_\mu \phi (D_\nu\phi)^\ast  -
g_{\mu\nu} (D_\alpha \phi)^\ast D^\alpha \phi - m^2 g_{\mu\nu}
|\phi|^2~, \nonumber\\
 \Theta_{\mu \nu}&=& \frac{2}{\sqrt{-g}}\frac{\delta\mathcal{L}_{\mathrm{int}}}{\delta g^{\mu\nu}}~.
\eea
The Maxwell equations are obtained  by varying the
Lagrangian with respect to
$A_\mu$ as
\be\label{maxeq}
\nabla_{\mu}F^{\mu \nu}= J^\nu + \mathcal{J}^\nu \ , \ee
where the current,
contains scalar and interaction  contributions, respectively,
\be J_\mu = i q \left[ \phi^\ast D_\mu
\phi -(D_\mu \phi)^\ast \phi\right] ~, \ \
\mathcal{J}_\mu = \frac{1}{\sqrt{-g}}\frac{\delta\mathcal{L}_{\mathrm{int}}}{\delta A^\mu}  ~. \ee Finally, varying the Lagrangian with respect to the
scalar field gives the scalar equation of motion as
 \bea D_\mu D^\mu \phi-m^2 \phi=\eta  D_\mu
\left( \mathcal{G}^{\mu\nu}D_\nu \phi\right)+\eta' D_\rho
(\mathcal{G}^{\mu\rho}D_\mu (D_\nu (\mathcal{G}^{\nu\sigma}
D_\sigma \phi)))~.
 \label{waveeq}
\eea
To capture the lattice effects, we consider the following metric
\textit{ansatz} \be
\begin{split} ds^2=&\frac{1}{z^2}\Big{[}-h(z)
Q_{tt}(x,z)dt^2+\frac{Q_{zz}(x,z)dz^2}{h(z)}\\&
 \ \ \ \ \ +Q_{xx}(x,z)(dx+z^2Q_{xz}(x,z)dz)^2+Q_{yy}(x,z)dy^2\Big{]}~,
\label{metriclat}
\end{split}
\ee
 where $h(z)$ is a fixed function, conveniently separated from the rest of $g_{tt}$, defined as
\begin{equation}
h(z)=1- \left( 1 + \frac{\mu_0^2}{4} \right) z^3+\frac{\mu_0^2}{4}z^4~.
\end{equation}
The required boundary conditions at the horizon and boundary
are, respectively,
\begin{equation}\label{QhorBC}
Q_{tt}(x,1)=Q_{zz}(x,1)~,
\end{equation}
and
\begin{equation}\label{QBbc}
Q_{tt}(x,0)=Q_{zz}(x,0)=Q_{xx}(x,0)=Q_{yy}(x,0)=1~,
~Q_{xz}(x,0)=0~,
~A_{t}(x,0)=\mu~,
\end{equation}
where we consider the system held under constant chemical potential $\mu$.

Notice that we have chosen coordinates in which the horizon is fixed at $z=1$. This fixes the overall scale of the system arbitrarily. Thus, the results ought to be reported in the form of dimensionless quantities. For example. the dimensionless temperature is given by
\begin{equation}
\frac{T}{\mu}=\frac{12-\mu_0^2}{16 \pi \mu}~,
\end{equation}
where $\mu = \mu_0$ above the critical temperature (in the absence of condensation of the scalar field). Below the critical temperature, $\mu$, measured in units of the radius of the horizon, increases.

The 
 solutions of the equations of motion have the same
form as found in \cite{Alsup:2013kda} at or above the critical temperature. In particular, the scalar field right below the critical temperature is not homogeneous, but of the form
\be
\phi(x, z)=\frac{\left<\mathcal{O}\right>}{\sqrt{2}}z^\Delta F(z)
\cos(kx) ~,\ \ F(0) =1\label{Fbc}\ee
leading to the formation of a one-dimensional lattice, where $\Delta$ is the
scaling dimension of the dual boundary operator $\mathcal{O}$.

Next, we will
 solve the equations of motion below the critical temperature  with the {\textit{ansatz}} \eqref{metriclat}.
To this end, we expand all the fields in the order
parameter
\begin{equation}
\xi = \frac{\langle \mathcal{O}\rangle}{\sqrt{2}}~,
\end{equation}
as \bea
Q_{tt}(x, z)&=&1+ \xi^2 Q_{tt}^1(x, z) + \mathcal{O} (\xi^4)~,\nonumber \\
Q_{zz}(x, z)&=& 1+\xi^2 Q_{zz}^1(x, z) + \mathcal{O} (\xi^4)~, \nonumber \\
Q_{xx}(x, z)&=&1+\xi^2 Q_{xx}^1(x, z) + \mathcal{O} (\xi^4)~, \nonumber \\
Q_{xz}(x, z)&=&\xi^2 Q_{xz}^1(x, z) + \mathcal{O} (\xi^4)~, \nonumber \\
Q_{yy}(x, z)&=&1+\xi^2 Q_{yy}^1(x, z) + \mathcal{O} (\xi^4)~, \nonumber \\
\phi(x, z)&=&\xi \phi^0(x, z)+\xi^3 \phi^1(x, z) + \mathcal{O}
(\xi^5)~, \nonumber
\\A_t(x, z)&=&(1-z)\left[ A_{t}^0(z)+\xi^2 A_{t}^1(z,x) +\mathcal{O}(\xi^4)\right]~,
\label{perturbation} \eea where
 $\phi^{0}$, and $A_{t}^{0}$ are  defined
at the critical temperature $T_c$.
The chemical potential (in units of the horizon radius) is given by \be\label{eq2_23} \mu \equiv
A_t (x, 0) = \mu_0 + \xi^2 \mu_1 + \mathcal{O} (\xi^2) \ \ , \ \ \
\ \mu_0 = A_{t}^0(0) \ \ ,  \ \ \ \ \mu_1 = A_{t}^1(x, 0)~. \ee
It should be emphasized that $\mu$ is a constant (independent of $x$), i.e., we impose the boundary condition that $A_t (x, 0)$ is a (fixed) constant. Dependence of the system on $x$ will be generated spontaneously.

The perturbative fields may be expanded in Fourier modes.
At each given order of the parameter $\xi$, only a finite number
of modes of the various fields are generated. At first order, i.e., $\mathcal{O}
(\xi^2)$, we have only the $0$ and $2k$ Fourier modes for the metric and gauge field functions, and $k$ and $3k$ modes for the scalar field,
\bea
Q_{tt}^1(x, z)&=& Q_{tt}^{1,0}(z)+ Q_{tt}^{1,1}(z) \cos 2kx ~,\nonumber \\
Q_{zz}^1(x, z)&=& Q_{zz}^{1,0}(z)+ Q_{zz}^{1,1}(z) \cos 2kx ~,\nonumber \\
Q_{xx}^1(x, z)&=& Q_{xx}^{1,0}(z)+ Q_{xx}^{1,1}(z) \cos 2kx ~,\nonumber \\
Q_{xz}^1(x, z)&=& Q_{xz}^{1,0}(z)+ Q_{xz}^{1,1}(z) \sin 2kx ~,\nonumber \\
Q_{yy}^1(x, z)&=& Q_{yy}^{1,0}(z)+ Q_{yy}^{1,1}(z) \cos 2kx ~,\nonumber \\
A_{t}^1(x, z)&=& A_{t}^{1,0}(z)+ A_{t}^{1,1}(z) \cos 2kx~,\nonumber \\
\phi^1(x,z) &=& \phi^{1,0}(z) \cos kx+\phi^{1,1}(z) \cos 3kx~. \label{fourier}
\eea
These modes can be obtained by solving the system of Einstein-Maxwell-scalar field equations. Details can be found in Appendix \ref{cmf}.

\begin{figure}[t]
\begin{center}
\includegraphics[width=.4\textwidth]{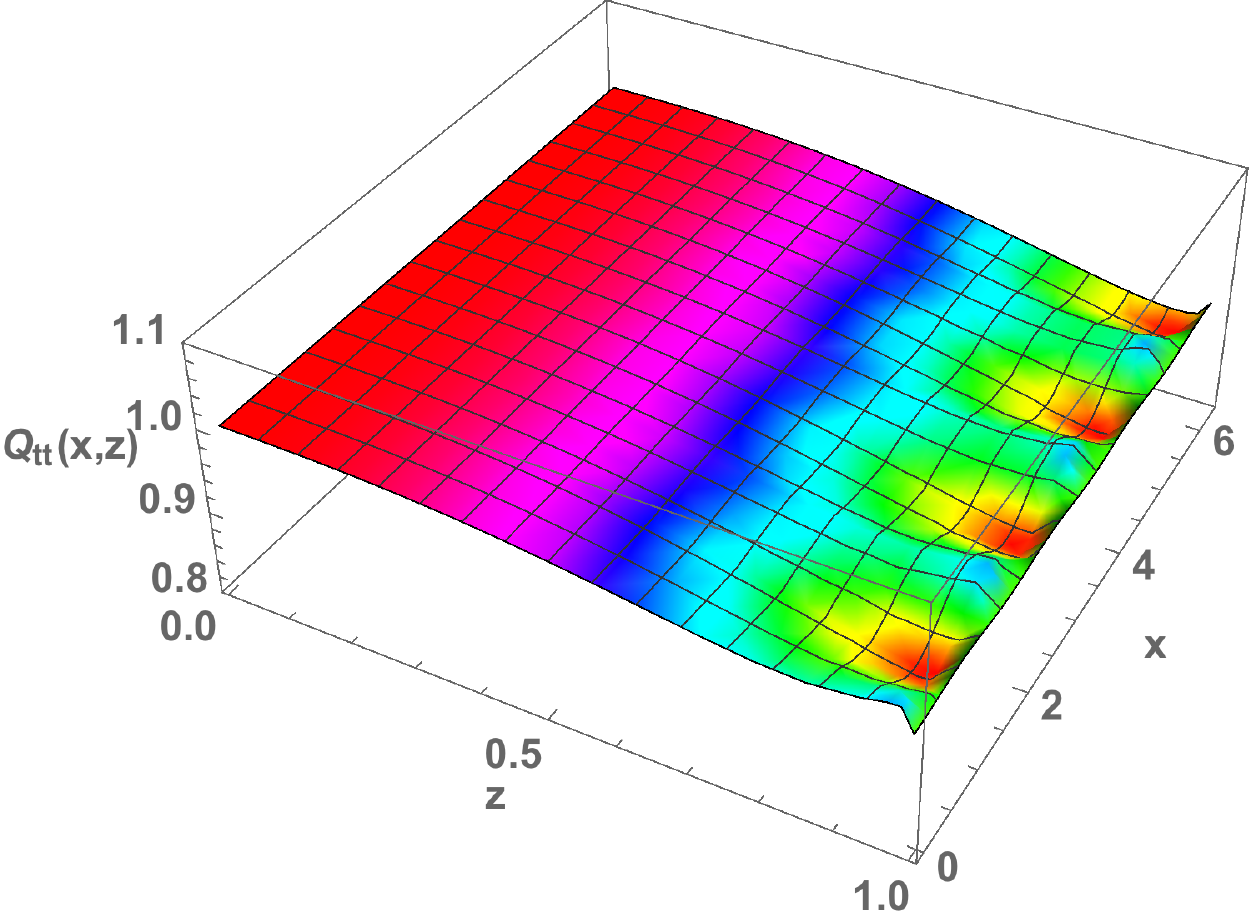}
\includegraphics[width=.4\textwidth]{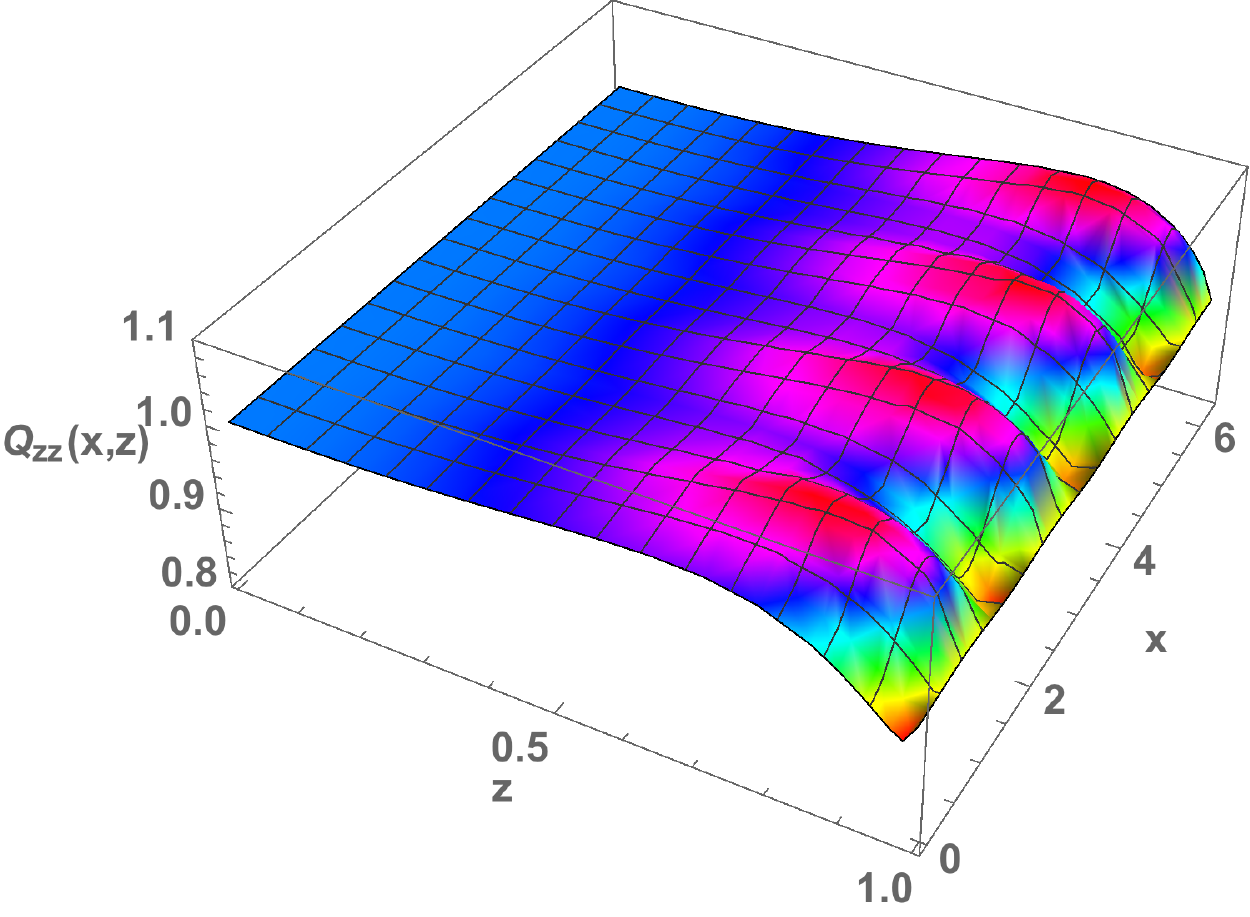}
\includegraphics[width=.4\textwidth]{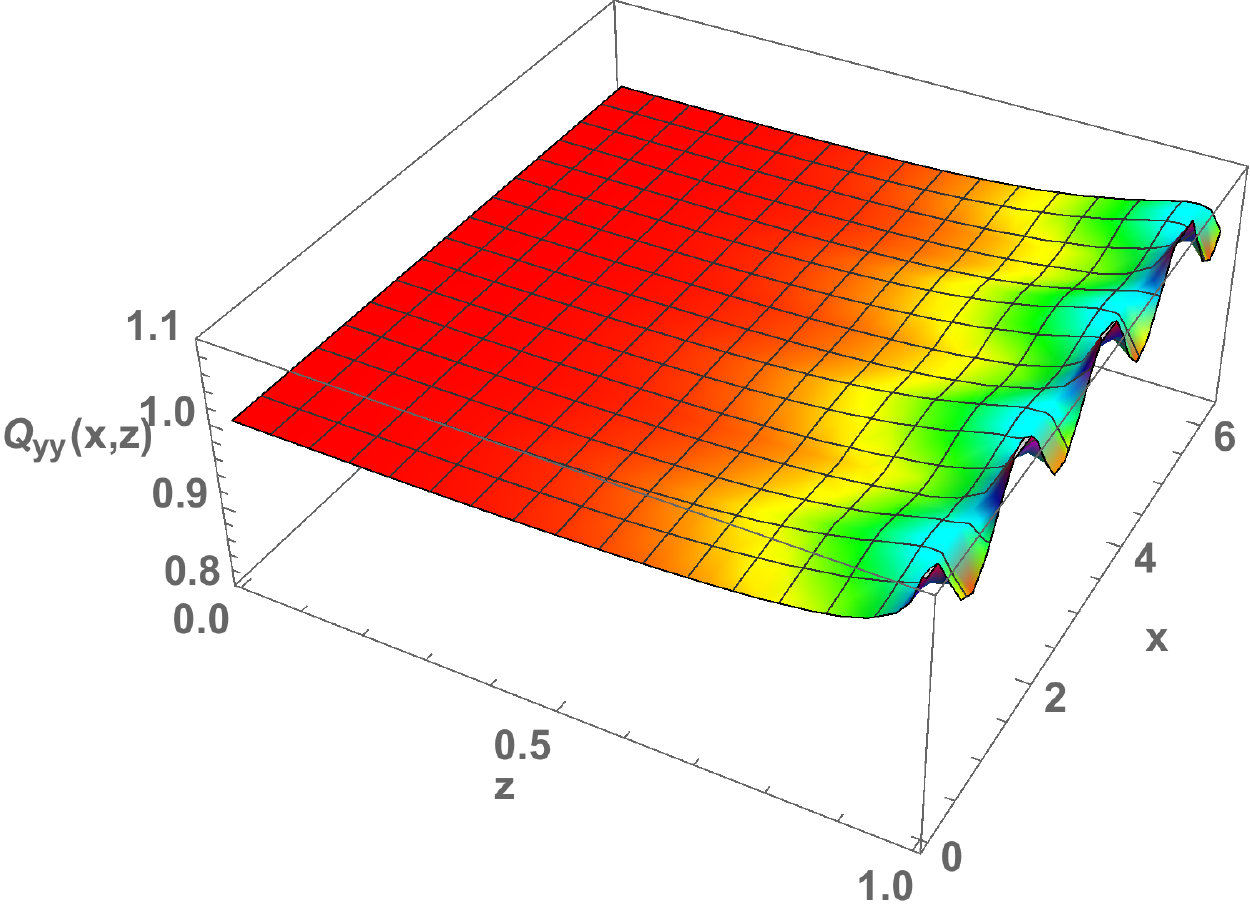}
\includegraphics[width=.4\textwidth]{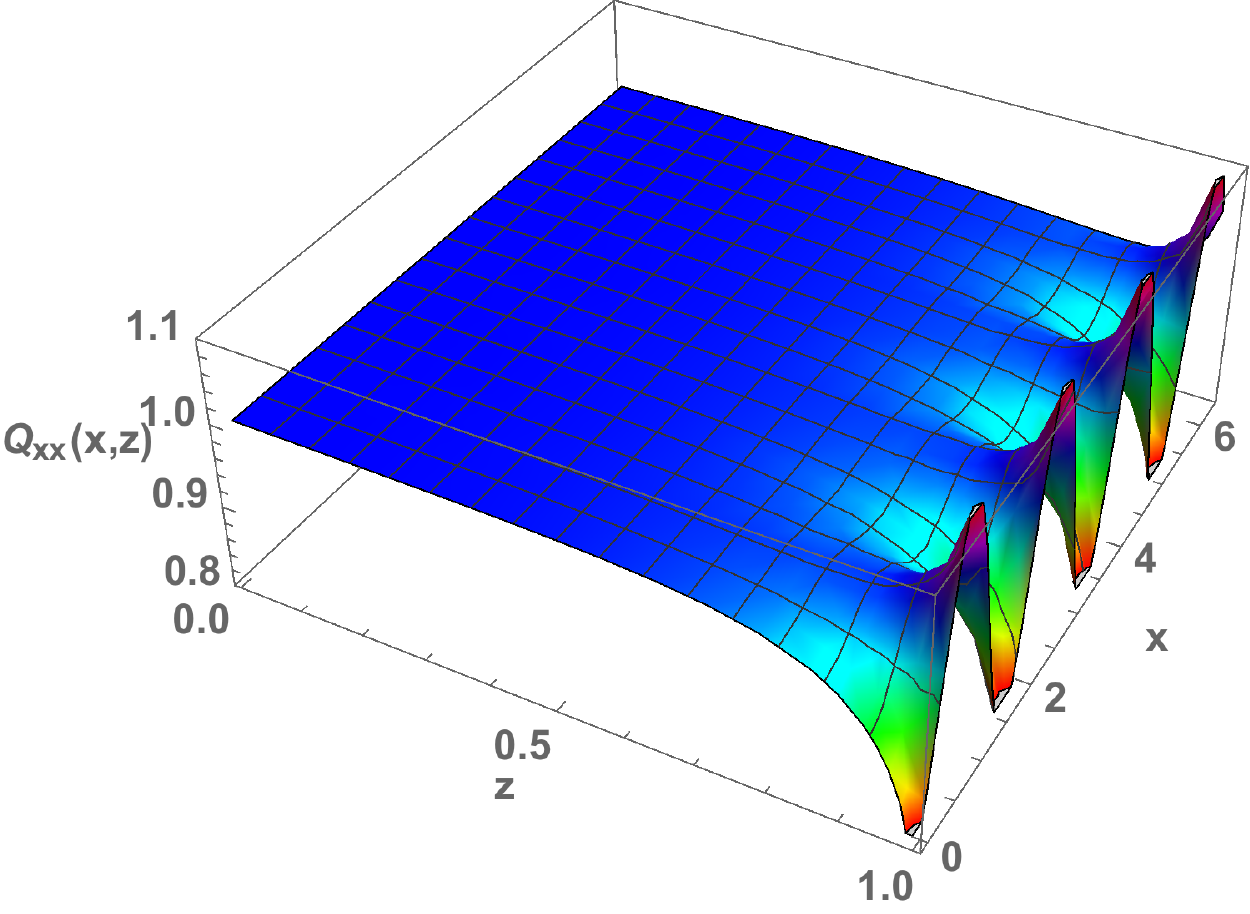}
\caption{Numerical solutions for the metric functions, $Q_{tt}(x, z) $, $Q_{zz}(x, z)$, $Q_{xx}(x, z)$, and
$Q_{yy}(x, z)$, for
$\frac{\eta}{\mu^2}=0.41$, $\frac{\eta'}{\mu^4}=0.005$, $q=0$,
$\Delta=1$. } \label{metricsolutions}
\end{center}
\end{figure}

\begin{figure}[t]
\begin{center}
\includegraphics[width=.5\textwidth]{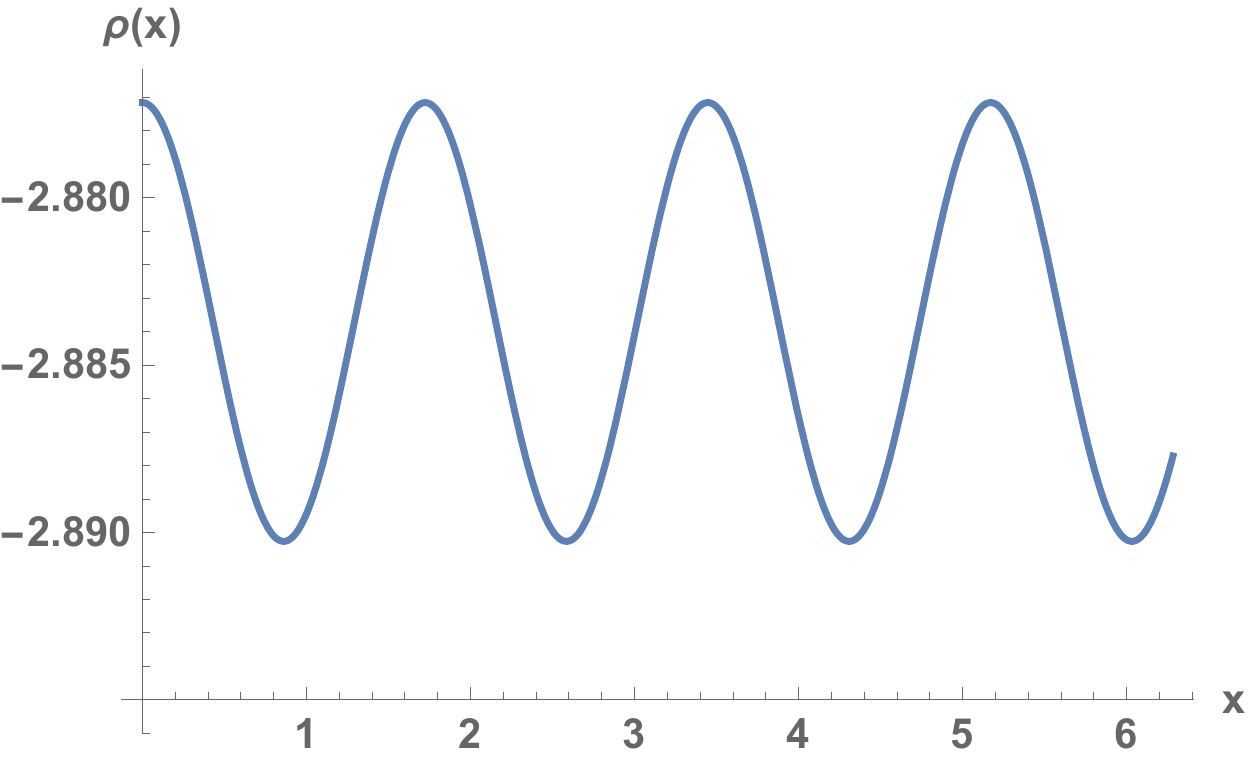}
\caption{The charge density of the system for
$\frac{\eta}{\mu_0^2}=0.41$, $\frac{\eta'}{\mu_0^4}=0.005$, $q=0$,
$\Delta=1$, and $\xi=0.1$. } \label{chargedensity}
\end{center}
\end{figure}

\begin{figure}[t]
\begin{center}
\includegraphics[width=.5\textwidth]{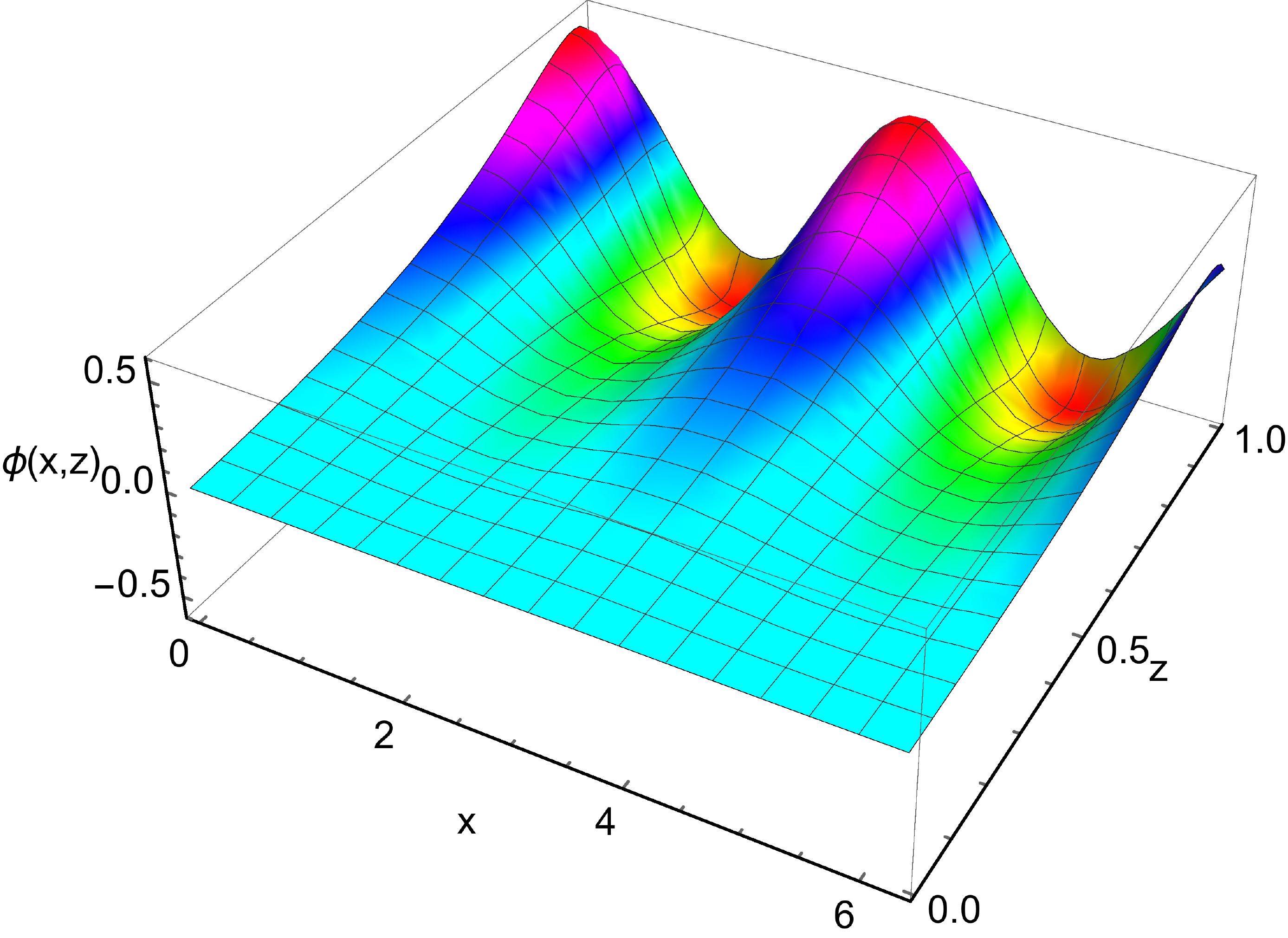}
\caption{The scalar field of the system with first order contributions included for parameters
$\frac{\eta}{\mu_0^2}=0.41$, $\frac{\eta'}{\mu_0^4}=0.005$, $q=0$,
$\Delta=1$, and $\xi=0.1$. } \label{scalarfirst}
\end{center}
\end{figure}

The charge density of the system can be determined through the dimensionless quantity
\begin{equation}
\frac{\rho}{\mu^2}=-\frac{\partial_z \left[(1-z) A_t(x, z)\right]|_{z=0}}{[A_t(x, 0)]^2}~.
\end{equation}
In Fig.\ \ref{metricsolutions}, we plot the modes of the metric
functions, $Q_{tt}(x, z)$, $Q_{zz}(x, z)$, $Q_{xx}(x,z)$, and $Q_{yy}(x, z)$, while the charge density of the system is
shown in  Fig.\ \ref{chargedensity}. As can be seen in these
figures, non-trivial $x-$spatially anisotropic  profiles of the
fields are developing below the critical temperature. The same behavior is also observed in the profile of the
scalar field, shown in Fig.\ \ref{scalarfirst}. In Fig.\ \ref{chargedensity}, the charge density is shown to be spatially
modulated, indicating a holographic lattice
structure below $T_c$. We observed that while keeping $\xi$ fixed and varying the other parameters, the spatial modulation varied as well. For example, as the critical temperature $T_c$ was lowered,
the magnitude of the spatial modulation decreased.

In summary, we have a perturbative, backreacted solution up to
first order in $\xi^2$ for the Einstein-Maxwell-scalar equations. Next-to-leading-order effects can be systematically introduced, but will not be needed for our purposes. Our focus is the behavior of the fermions and next-to-leading-order corrections to the other fields contribute negligibly to the leading order behavior of the fermionic fields.

\section{The Dirac Equation}
\label{Diracsection}

In this section we  introduce  a Dirac spinor field
$\Psi(z,t,x,y)$ with mass $m_f$ and charge $q_f$, in addition
to the scalar, gauge, and gravitational fields introduced in the previous section. The bulk action for the Dirac field is given by
\begin{equation}
    S_D=i \int d^4 x \sqrt{-g} {\bar{\Psi}} (\Gamma^a \mathcal{D}_a
    -m_f)\Psi~, \label{diraction}
\end{equation}
where $\Gamma^a=(e_\mu)^a \Gamma^\mu$ with a set of orthogonal normal vector bases $(e_\mu)^a$ and $\Gamma^\mu$ the Dirac gamma matrices.
The covariant derivative $\mathcal{D}_a$ is defined as
\begin{equation}
    \mathcal{D}_a=\partial_a+\frac{1}{4}(\omega_{\mu\nu})_a \Gamma^{\mu\nu}-iq_f A_a~,
\end{equation}
with $\Gamma^{\mu\nu}=\frac{1}{2}[\Gamma^\mu, \Gamma^\nu]$, and
$(\omega_{\mu\nu})_a=(e_\mu)_b \nabla_a (e_\nu)^b $ are the spin
connection 1-forms. The indices $a$, $b$ denote tangent space, and the $\mu, \nu$
indices  denote the boundary directions. The gamma
matrices satisfy the Clifford algebra
$\{\Gamma^a, \Gamma^b\}=2 \eta^{ab}$, and $\Gamma_{ab}=\frac{1}{2}[\Gamma_a, \Gamma_b ]$.

From the action (\ref{diraction}) with the  metric
(\ref{metriclat}), the Dirac equation is of the form
\begin{equation}
    \begin{split}
        &-\sqrt{\frac{h}{Q_{zz}}}\Gamma^4 \partial_z\Psi + \frac{1}{\sqrt{hQ_{tt}}}
        \Gamma^1 \left(\partial_t-iq_f A_t\right)\Psi+\frac{1}{\sqrt{Q_{xx}}}
        \Gamma^2 \partial_x \Psi+\frac{1}{\sqrt{Q_{yy}}} \Gamma^3 \partial_y
        \Psi-\frac{m_f}{z} \Psi\\&+\frac{1}{4\sqrt{Q_{xx}}}  \partial_x \ln \frac{Q_{tt} Q_{zz} Q_{yy}}{z^6}\Gamma^2 \Psi -\frac{1}{4}\sqrt{\frac{h}{Q_{zz}}}
        \partial_z \ln \frac{hQ_{tt} Q_{xx} Q_{yy}}{z^6} \Gamma^4
        \Psi=0~.
    \end{split} \label{direqu}
\end{equation}
We choose the basis
\begin{equation}\label{gammaEqn}
    \Gamma^1=\begin{pmatrix} i \sigma^1&0\\0& i\sigma^1 \end{pmatrix}, \ \ \ \Gamma^2=\begin{pmatrix} -\sigma^2&0\\
     0&\sigma^2\end{pmatrix}, \ \ \ \Gamma^3=\begin{pmatrix} 0&\sigma^2\\ \sigma^2&0\end{pmatrix}, \ \
     \ \Gamma^4=\begin{pmatrix} -\sigma^3&0\\0& -\sigma^3\end{pmatrix}
\end{equation}
for the gamma matrices of the (3+1)-dimensional bulk theory,
and the following {\textit{ansatz}} for the spinor fields
\begin{equation}
    \Psi= \left( \frac{hQ_{tt} Q_{xx} Q_{yy}}{z^6} \right)^{-\frac{1}{4}} e^{-i \omega t+i (k_x x+k_y y)}\psi \ \ , \ \ \ \ \psi = \begin{pmatrix}
    \psi_+\\ \psi_- \end{pmatrix} \ \  , \ \ \ \ \psi_\pm=\begin{pmatrix} \psi_{\pm 1}\\ \psi_{\pm 2}
    \end{pmatrix}~.\label{spinor1}
\end{equation}
%
We have a holographic lattice structure in the $x$ dimension.
Therefore, we can expand the spinor fields, according to the Bloch
theorem, as
\begin{equation}
    \psi_{\alpha s}(x,z) =\sum_{l=0,\pm 1,\pm 2, \dots}
   \psi_{\alpha s}^{l}(z) e^{2i l k x}~, \label{Bloch}
\end{equation}
where $\alpha = \pm, s=1,2$.

Upon substituting (\ref{Bloch}) into the Dirac equation
(\ref{direqu}), we obtain a set of coupled equations with
infinitely many fields, corresponding to the full range
of $l = 0, \pm 1, \pm 2, \dots, \pm \infty$. Our aim is to calculate the retarded Green function and
 the spectral function to characterize the fermionic system. This is
equivalent to measuring the spectral function by   Angular
Resolved Photoemission Spectroscopy (ARPES),  as  discussed in \cite{Faulkner:2010da}. To this end,
we will numerically and analytically solve the equations using
perturbation theory at small but finite temperature, and compare
the behavior of these solutions to those above the critical temperature.


The backreaction contribution of the scalar field, gauge field, and metric are of
first order in $\xi$ for our expansion \eqref{perturbation}.  To solve the
Dirac equation for the spinor field, we expand in
$\xi$ up to second order as
\be
\psi_{\alpha s}^{l}(z) = \psi_{\alpha s}^{0, l} +\xi^2  \psi_{\alpha s}^{1, l}+\xi^4  \psi_{\alpha s}^{2, l}+\mathcal{O}(\xi^6)~. \label{pertDirac}
\ee
Near the horizon, we have ingoing boundary conditions,
\begin{equation}
\begin{pmatrix} \psi_{\pm 1}\\ \psi_{\pm 2} \end{pmatrix}
\sim\begin{pmatrix} 1 \\ -i \end{pmatrix} (1-z)^{-\frac{i \omega}{4 \pi T}}~,
\end{equation}
To implement them, we write
\be \psi_{\alpha s} (z) = (1-z)^{-\frac{i \omega}{4 \pi T}} F_{\alpha s} (z)~, \ee
where $F_{\alpha 1} =1$ and $F_{\alpha 2} = -i$, at $z=1$.

For fixed $\alpha, l$, we obtain a unique solution by setting $\psi_{\beta s}^{0, l'} = 0$, for all $(\beta ,l') \ne (\alpha, l)$. Physically, this means that at the critical temperature, we have a single mode labeled by $\alpha,l$. At higher orders in $\xi$, i.e., below the critical temperature, more modes are generated.

At the AdS boundary ($z \to 0$), the mode of the solution to the Dirac equation labeled by $\beta,l'$ asymptotes to
\begin{equation}
    \psi_{\beta }^{l'} (z) \approx A_{\alpha \beta}^{ll'}\,
     z^{-m_f} \begin{pmatrix} 0 \\ 1 \end{pmatrix}+B_{\alpha \beta }^{ll'}\, z^{m_f} \begin{pmatrix} 1 \\ 0 \end{pmatrix}~, \label{boundarydirac}
\end{equation}
where $\alpha, l$ encode the choice of solution at the critical temperature.
The retarded Green function can be found from the matrices formed by the coefficients in the asymptotic expansion, through the matrix
\begin{equation}
   \mathbf{G}_R = \mathbf{BA}^{-1}~.
    \label{retG}
\end{equation}
%
We will solve the Dirac equation up to second order in $\xi^2$.
To find the explicit form of the retarded Green function from (\ref{retG}), we note that the diagonal elements of the matrix $\mathbf{A}$ have contributions which are $\mathcal{O} (\xi^0)$ and $\mathcal{O}(\xi^4)$, whereas the off-diagonal elements are $\mathcal{O} (\xi^2)$. These follow directly from the Dirac equation. The retarded Green function for the solution labeled by $\alpha,l$ can therefore be written as
\be
G_R = \left( \mathbf{G}_{R} \right)_{\alpha\alpha}^{ll} =\frac{1}{\det A}\left\{ \sum_{\beta=+,-}\left(B_{\alpha \beta}^{l l-1} \Delta_{\alpha  \beta}^{l l-1}+B_{\alpha \beta}^{l l+1}\Delta_{\alpha  \beta}^{l l+1}\right)+B_{\alpha  \alpha}^{l l} \Delta_{\alpha  \alpha}^{l l} + \mathcal{O} (\xi^4)\right\} ~, \label{eqGR}
\ee
where $\Delta_{\alpha  \beta}^{l l'}$ are the cofactors of the matrix $A$~\cite{LSSZ:2012}.

The spectral function
is the imaginary part of the diagonal terms of the
retarded Green function,
\be\label{eq32} A^l (\omega, k_x, k_y)=\Im \left[ \left( G_R \right)_{+ }^l + \left( G_R \right)_{- }^l \right]~,\ee
and gives the location of the
Fermi surface. Although the Fermi surface is defined at zero temperature, as
indicated in \cite{Ling:2013fl} and \cite{BHY2010}, it
may be located by searching for a peak in the spectral function $A(\omega,
k_x, k_y)$ at small frequency $\omega$ and low temperatures.

\section{Fermionic solution at the critical temperature}
\label{DiracZeroth}

In this section we discuss the physics of the fermionic solution at small but non-zero critical temperature. Here and in subsequent sections, we focus on low frequency modes ($\omega \ll \mu_0$), because they determine the macroscopic properties of the system. Results are obtained both analytically and numerically in the low-temperature limit.
The Dirac equation at the critical temperature reads
\be
\partial_z {\psi}^{0, l}_{\alpha s}-i\frac{q_f \mu_0(1-z)+\omega}{h(z)}\sigma^2{\psi}^{0,l}_{\alpha s}+
\frac{k_x+2k l}{\sqrt{h(z)}}\sigma^3{\psi}^{0,l}_{\alpha s}-\frac{
	k_y}{\sqrt{h(z)}}\sigma^1 \psi^{0, l}_{-\alpha \ 3-s}=0~,
\label{DiracaboveTc} \ee where $\alpha=\pm$, $s=1, 2$, and
we have set
$m_f=0$.


To obtain the analytic solution at low frequencies, we solve the Dirac equation in the near-horizon region ($1-z\ll 1$) and in the far region ($1-z\gg \omega/\mu_0$), and then match the two solutions in the overlapping region ($\omega/\mu_0 \ll 1-z \ll 1$).

In the small temperature limit, the near-horizon
region  is AdS$_2\times \mathbb{R}^{2}$, and the metric
can be written as \cite{Faulkner:2009wj}
\begin{equation}
ds^2=\frac{1}{6u^2}\left(-(1-\frac{u^2}{u_0^2})dt^2+\frac{du^2}{1-\frac{u^2}{u_0^2}}\right)+ dx^2+ dy^2
\end{equation}
after changing coordinates to $u = \frac{\omega z_\ast^2}{6(z_\ast -z)}$, where $z_\ast = \frac{2\sqrt{3}}{\mu_0}$. The horizon is at $u_0 = \frac{\omega z_\ast^2}{6(z_\ast -1)}$ with corresponding temperature $T=1/2\pi u_0$,
and the $U(1)$ gauge field is given by
\be
A_{t}^0 =\frac{1}{\sqrt{3}}\left( \frac{1}{u}-\frac{1}{u_0}\right)~. \ee
After performing the $SO(2)$ rotation discussed in Appendix \ref{spinorrota}, it is convenient to express the rotated Dirac field in the near-horizon region as
\be\label{phitilde}
\tilde{\psi}_{\alpha s}^{0,l} =\frac{1}{\sqrt{2}}\left( 1-\frac{u^2}{u_0^2} \right)^{-1/4} (1+i \sigma^1)\tilde{y}_{\alpha s}^{0,l} (u)~.
\ee
The massless Dirac equation at zeroth order becomes independent of $\omega$,
\be\label{eq0Dirac}
\partial_u \tilde{y}_{\alpha s}^{0,
	l}	-\frac{i}{ 1 - \frac{u^2}{u_0^2}} \sigma^3\left( 1 + q_f
A_{t}^{0}\right)\tilde{y}_{\alpha s}^{0, l}-\frac{ \alpha \sqrt{2}k_l}{\mu_0u\sqrt{ 1 - \frac{u^2}{u_0^2}}
} \sigma^1 \tilde{y}_{\alpha s}^{0, l} 
\ee
where $k_l^2 = (k_x+2kl)^2 + k_y^2$.
For the upper component of $\tilde{y}_{\alpha s}^{0,l} (u)$,
we obtain the second-order equation
\be\label{secondordery} \mathbf{L}_2 [\tilde{y}_{\alpha 1}^{0, l}] = 0 \ee
where
\be
 \mathbf{L}_2 [\tilde{y}] \equiv  \partial_u^2 \tilde{y} +\frac{1 - 2 \frac{u^2}{u_0^2}}{u( 1 - \frac{u^2}{u_0^2})} \partial_u \tilde{y}
+\left[-\frac{\nu_{k_l}^2}{u^2 (1-\frac{u^2}{u_0^2})}+\frac{\left( 1+q_f A_t^0\right)^2 - \frac{q_f^2A_t^0 }{\sqrt{3}}\left( \frac{1}{u_0}+\frac{1}{u}\right)-\frac{i q_f A_t^0 }{ u_0}- \frac{i }{u}}{(1-\frac{ u^2}{u_0^2})^2} \right] \tilde{y}\label{secondordery2}
\ee
and we defined
\be \nu_{k_l}=\frac{\sqrt{2}}{\mu_0}\sqrt{k_l^2-\frac{q_f^2 \mu_0^2}{6}}~. \ee
The solution satisfying ingoing boundary conditions at the horizon may be expressed in terms of hypergeometric functions (up to an irrelevant normalization constant)
\bea\label{yInOut}
	\tilde{y}_{\alpha 1}^{0, l} &=& \left(1+\frac{u_0}{u}\right)^{\frac{1}{2}+\frac{i q_f}{\sqrt{3}} }
	 \left(\frac{u_0^2}{u^2}-1\right)^{-\frac{i u_0}{2}} \nonumber\\
	 && \times \ {_2}F_1 \left(\frac{1}{2}-\nu_{k_l}+\frac{i q_f}{\sqrt{3}} -i u_0,\frac{1}{2}+\nu_{k_l}+ \frac{i q_f}{\sqrt{3}} -i u_0;\frac{1}{2}-i u_0;\frac{u-u_0}{2u}\right)
~.\ \ \ \ \ \
\eea
Working similarly, we obtain the lower component of $\tilde{y}_\alpha^{0, l}$,
\bea\label{zInOut}
\tilde{y}_{\alpha 2}^{0, l} &=& \left(1+\frac{u_0}{u}\right)^{\frac{1}{2}-\frac{i q_f}{\sqrt{3}} +\frac{i u_0}{2}} \left(\frac{u_0}{u}-1\right)^{\frac{1}{2}-\frac{i u_0}{2}}
\nonumber\\ &&  \times \ {_2}F_1\left(1-\nu_{k_l}-\frac{i q_f}{\sqrt{3}} ,1+\nu_{k_l}-\frac{i q_f}{\sqrt{3}} ;\frac{3}{2}-i u_0;\frac{u-u_0}{2 u}\right)~.\ \ \ \ \ \
\eea
To match the above near-horizon solution with the solution in the far region, we need its asymptotic behavior away from the horizon ($u\to 0$). Using the properties of hypergeometric functions, we obtain, after switching coordinates, an asymptotic expression of the form
\be
\tilde{y}_\alpha^{0,l}\sim \begin{pmatrix} - \nu_{k_l} \\ \alpha \frac{\sqrt{2} k_l}{\mu_0}+ \frac{q_f}{\sqrt{3}} \end{pmatrix} (z_\ast -z)^{\nu_{k_l}}+ \mathcal{G}_R(\omega)   \begin{pmatrix} + \nu_{k_l} \\ \alpha \frac{\sqrt{2} k_l}{\mu_0}+ \frac{q_f}{\sqrt{3}} \end{pmatrix} (z_\ast -z)^{-\nu_{k_l}}
\ee
where $\mathcal{G}_R(\omega)$ is found explicitly.
By matching the above expression with the solution in the far region in the overlap region, after some algebra (for details, see ref.\ \cite{Faulkner:2009wj}), we arrive at the retarded Green function at the critical temperature,
\be
\mathcal G_R=(4 \pi T_c)^{2\nu_{k_l}}\frac{-\frac{i\sqrt{2}k_l}{\mu_0} +\frac{i
	q_f}{\sqrt{3}} +\nu_{k_l}}{-\frac{i\sqrt{2} k_l}{\mu_0} +\frac{i q_f}{\sqrt{3}}
	-\nu_{k_l}}\cdot\frac{\Gamma (-2 \nu_{k_l}) \Gamma (1+\nu_{k_l}-\frac{i
	q_f}{\sqrt{3}} ) \Gamma (\frac{1}{2}+\frac{i q_f}{\sqrt{3}} +\nu_{k_l}-\frac{i
		\omega}{2 T_c} )}{\Gamma (2 \nu_{k_l}) \Gamma (1-\frac{i q_f}{\sqrt{3}}
	-\nu_{k_l}) \Gamma (\frac{1}{2}+\frac{i q_f}{\sqrt{3}} -\nu_{k_l}-\frac{i
		\omega}{2 T_c} )}~. \label{GreensAna} \ee
The form of \eqref{GreensAna} controls the shape of the full Green function's poles at $k_F$.


Next, we solve the Dirac equation numerically and compare with the expectations from the analytical work.
After obtaining the numerical solution, the spectral function is found by evaluating the boundary behavior via
\be A^l (\omega, k_x, k_y)=\Im\left[\frac{\psi_{+ 1}^{0, l}(\epsilon)}{\psi_{+ 2}^{0, l}(\epsilon)}+\frac{\psi_{- 1}^{0, l}(\epsilon)}{\psi_{- 2}^{0, l}(\epsilon)}\right]~ \label{specfunc1}\ee
\begin{figure}[t]
	\begin{center}
		\includegraphics[width=.7\textwidth]{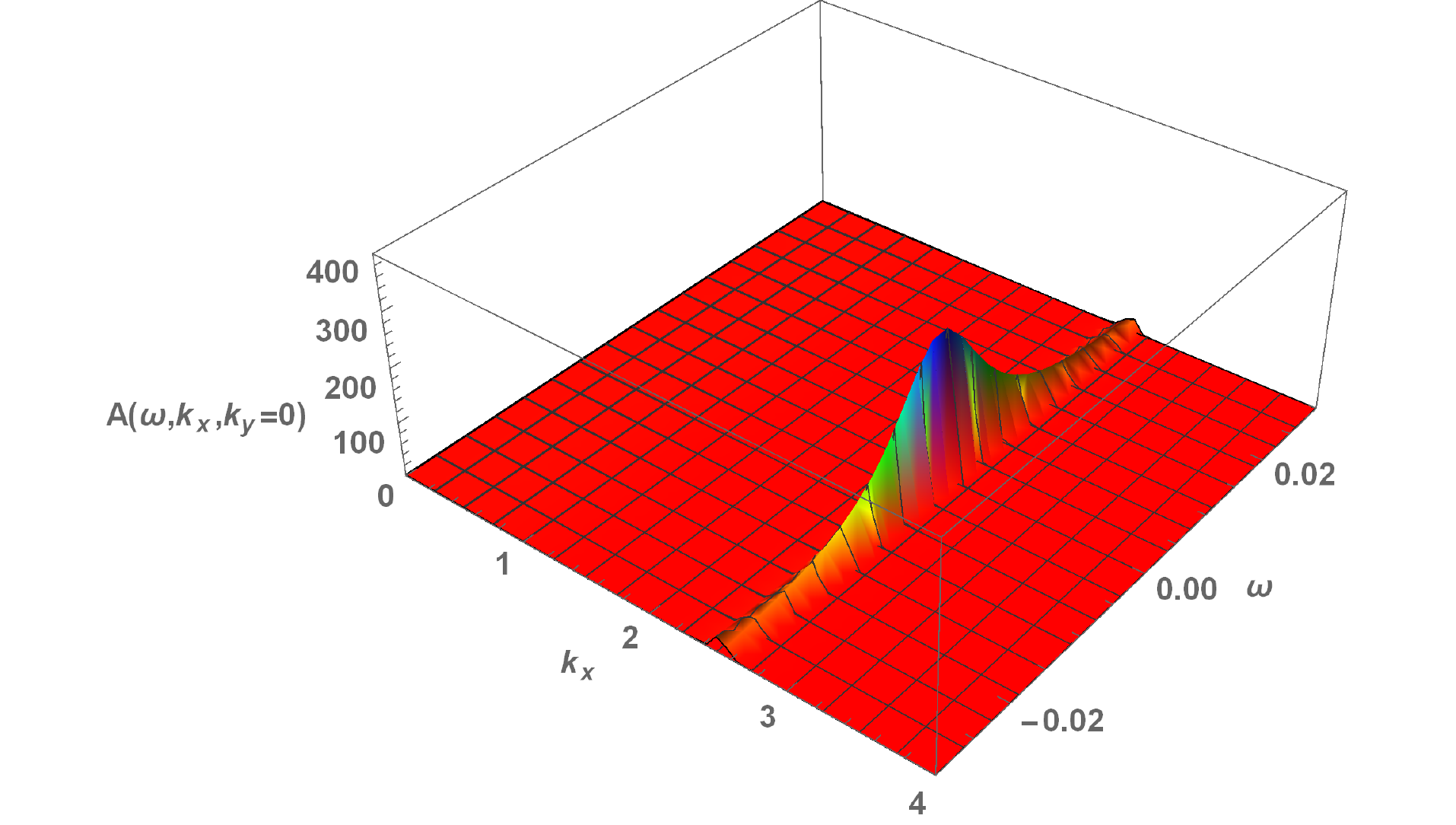}
		\caption{The spectral function $A^0(\omega, k_x, k_y=0)$ at the critical temperature calculated numerically, showing a peak at $k_F=2.573$, for $q_f=1.7$, $\mu_0=2.35$,
			and scalar $\Delta=1$. }
		\label{specfunc}
	\end{center}
\end{figure}
in the limit $\epsilon\to 0$. Using \eqref{specfunc1},
the
spectral function at the critical temperature $T_c$ and small frequencies is plotted
in Fig.\ \ref{specfunc} for different values of $k_x$ and $k_y=0$. The plot shows the Fermi surface as a peak of the spectral function
at $k_x = k_F=2.653$. However, the
peak is quite broad indicating an instability of the
quasi-particles.

\begin{figure}[t]
	\begin{center}
		\includegraphics[width=.6\textwidth]{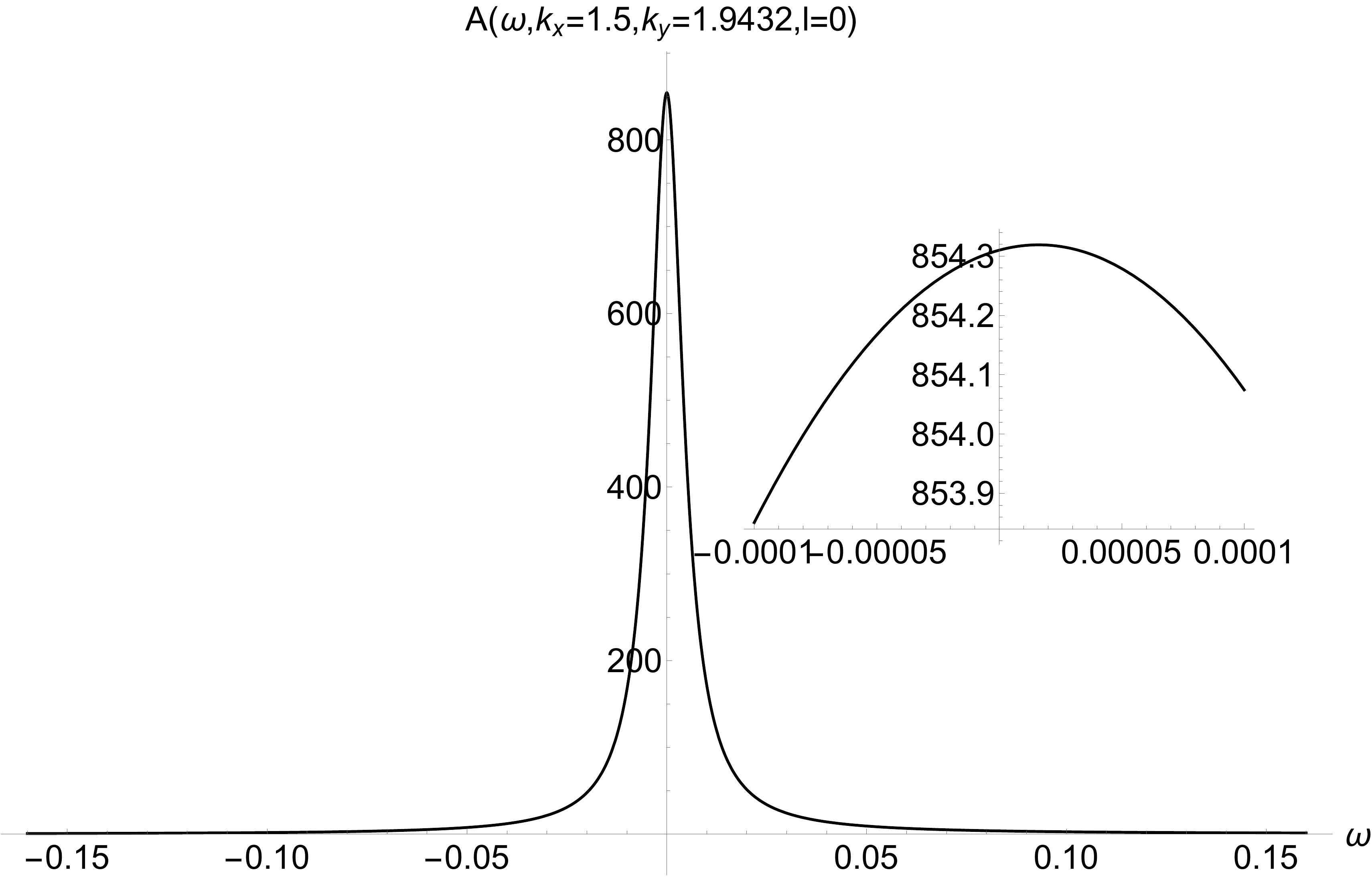}
		\caption{The spectral function $A^0(\omega, k_x=1.5, k_y=1.9432)$ at the critical temperature calculated numerically, showing a peak at $k_F=\sqrt{k_x^2 + k_y^2} = 2.4548$, for $q_f=1.7$, $\mu_0=2.2507$, and scalar $\Delta=1$. }
		\label{kx1ky2}
	\end{center}
\end{figure}
\begin{figure}[t]
	\begin{center}
		\includegraphics[width=.6\textwidth]{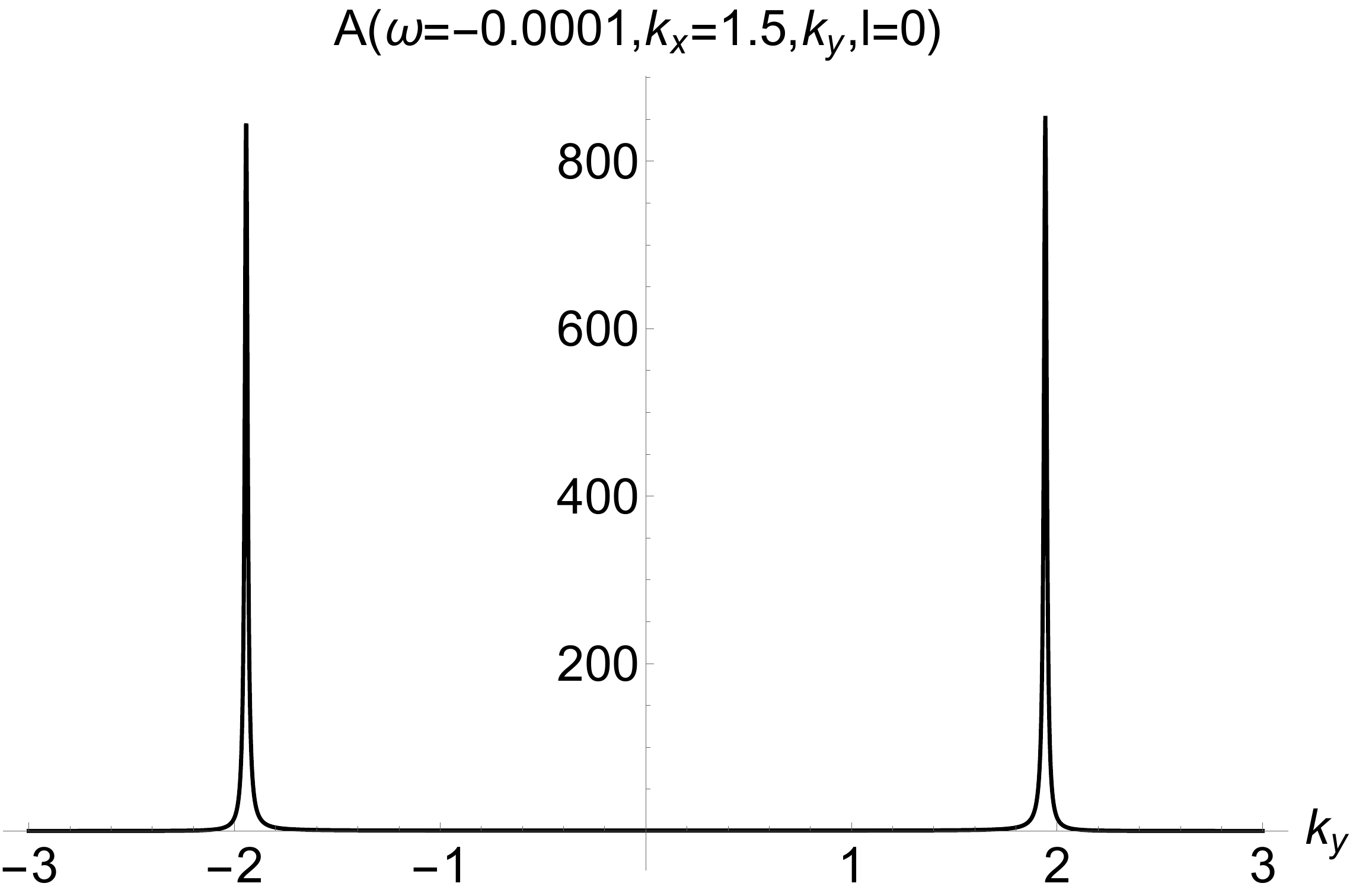}
		\caption{The spectral function $A^0(\omega=-0.0001, k_x=1.5, k_y)$ at the critical temperature calculated numerically, showing peaks at $k_y = \pm 1.9432 $, implying $k_F=\sqrt{k_x^2+k_y^2}=2.4548$, for $q_f=1.7$, $\mu_0=2.2507$, and scalar $\Delta=1$. }
		\label{kx1ky2k}
	\end{center}
\end{figure}

\begin{figure}[t]
	\begin{center}
		\includegraphics[width=.6\textwidth]{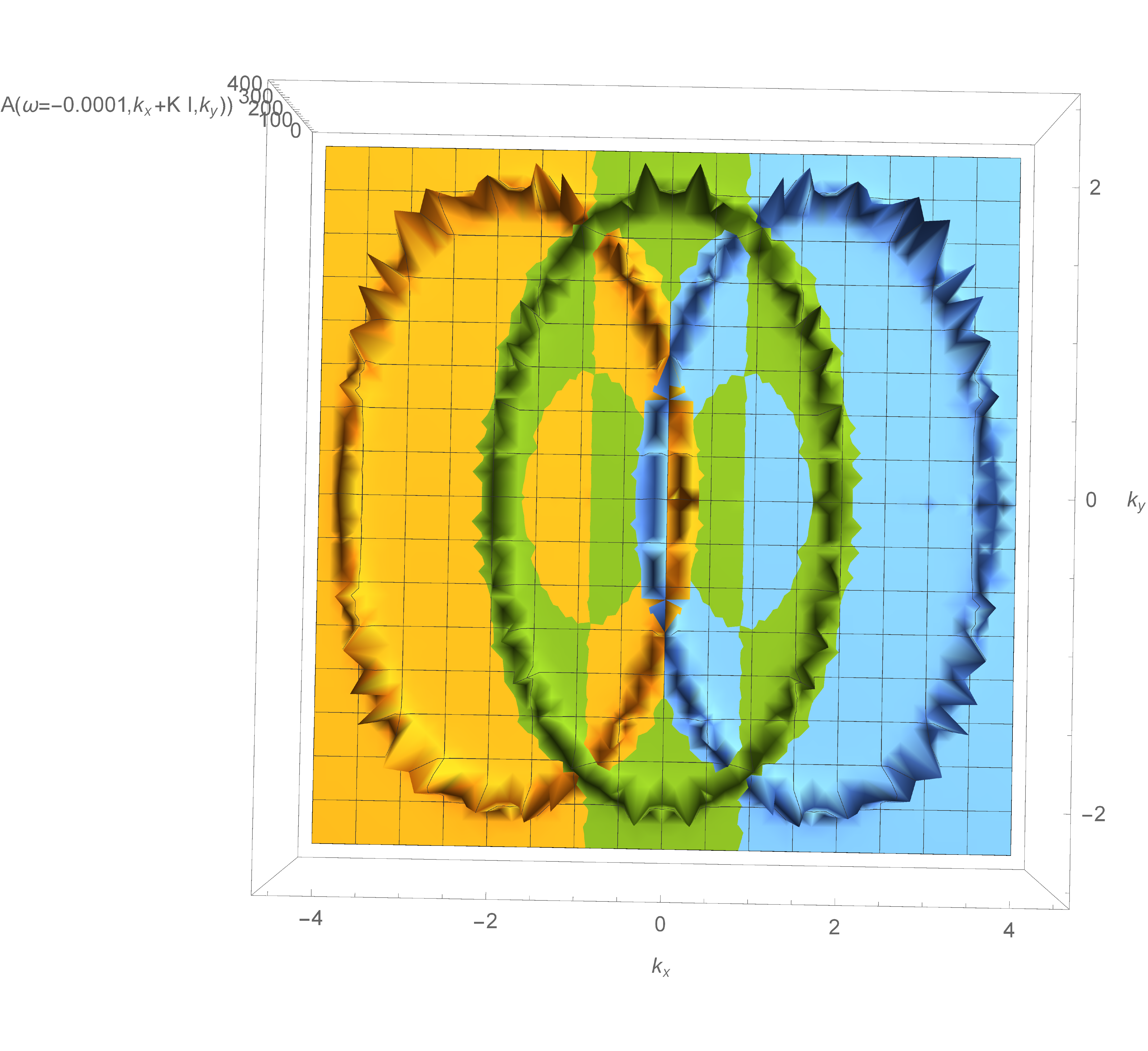}
		\caption{The spectral function
			$A^l(\omega=-0.0001, k_x, k_y)$ at the critical temperature calculated numerically for $l=-1,0,+1$, $q_f=1.56$, $\mu_0=2.2507$,
			and scalar $\Delta=1$. } \label{l1l0lm1}
	\end{center}
\end{figure}

Using the rotation outlined in Appendix \ref{spinorrota}, we calculate numerically the spectral function at the critical temperature for
non-zero values of both $k_x$ and $k_y$ and plot the results in Figures \ref{kx1ky2},
\ref{kx1ky2k}, and \ref{l1l0lm1}. Figure \ref{kx1ky2} is a plot of
the spectral function with $k_x=1.5, k_y=1.9432$ as a function of
$\omega$ at the location of the peak along the line $k_y =1.9432$. There is a peak at small frequency with
$k_F=\sqrt{k_x^2+k_y^2}=2.4548$ for parameters $q_f=1.7$, $\mu_0=2.2507$,  and scalar dimension $\Delta=1$
at the critical temperature. Similarly, we plotted the
spectral function as a function of $k_y$ in Fig.
\ref{kx1ky2k} for the same parameters and at small frequency ($\omega = -.0001 $). The
plot shows a peak at $k_y =\pm 1.9432$ confirming that
$k_F=2.4548$. 
Finally, Fig. \ref{l1l0lm1} shows
the Fermi surface  for $l=1, 0, -1$, from left to
right, respectively. We used the same parameters for plotting Fig. \ref{l1l0lm1} but the fermion charge is $q_f=1.56$. As seen in Fig. \ref{l1l0lm1} the $k_F$ value is smaller for a smaller $q_F$ value. 

\section{Fermionic solution below the critical temperature}
\label{DiracFirst}

Having obtained the solution to the Dirac equation at the small critical temperature, we proceed to calculate the solution below the critical temperature perturbatively. For non-trivial physical results, it is necessary to include second-order effects.

\subsection{First Order}
\label{DiracFirstAna}

Starting with the $l$-th mode at the critical temperature, at $\mathcal{O} (\xi^2)$ below the critical temperature, three modes are excited, with $l' = l, l\pm 1$.
For an analytic solution, as in Section \ref{DiracZeroth}, we need to analyze the near-horizon region. After performing the rotation described in Appendix \ref{spinorrota} at first order,
\be \left( \begin{array}{c} \tilde{\psi}_{+s}^{1,l} \\ \tilde{\psi}_{-s}^{1,l} \end{array} \right) = \left( \begin{array}{cc} \cos\frac{\theta}{2} & - \sin\frac{\theta}{2} \\ \sin\frac{\theta}{2} & \cos\frac{\theta}{2} \end{array} \right) \left( \begin{array}{c} {\psi}_{+s}^{1,l} \\ {\psi}_{-s}^{1,l} \end{array} \right) \ee
where $\tan\theta = \frac{k_y}{k_x+2kl}$, and defining
\be\label{phitilde1}
\tilde{\psi}_{\alpha s}^{1,l} =\frac{1}{\sqrt{2}}\left( 1-\frac{u^2}{u_0^2} \right)^{-1/4} (1+i \sigma^1)\tilde{y}_{\alpha s}^{1,l} (u)~.
\ee
the first-order massless Dirac equation near the horizon becomes
\be\label{eq1Dirac}
\partial_u \tilde{y}_{\alpha s}^{1,
	l'}	-\frac{i}{ 1 - \frac{u^2}{u_0^2}} \sigma^3\left( 1 + q_f
A_{t}^{0}\right)\tilde{y}_{\alpha s}^{1, l'}-\frac{ \alpha \sqrt{2}k_l}{\mu_0u\sqrt{ 1 - \frac{u^2}{u_0^2}}
} \sigma^1 \tilde{y}_{\alpha s}^{1, l'} + \mathcal{A}_{\alpha s}^{l'}=0~,
\ee
where
\be
\begin{split}
\mathcal{A}_{\alpha s}^{l} =-&
	\frac{1}{2}Q_{zz}^{1,0}\partial_u \tilde{y}^{0, l}_{\alpha s} +\frac{i\left[\frac{q_f}{\sqrt{3}} \left(\frac{u_0}{u}-1\right)Q_{tt}^{1,0}+2 q_f (u-u_0)A_{t}^{1,0}\right]}{2 u_0 \left(1 - \frac{u^2}{u_0^2}\right)}\sigma^3 \tilde{y}^{0, l}_{\alpha s}
\\&-\alpha \frac{\left[(k_x+2k l)^2Q_{xx}^{1,0}-\left((k_x+2kl)^2-k_l^2\right)Q_{yy}^{1,0}\right]}{\sqrt{2}\mu_0 k_{l} u\sqrt{1 - \frac{u^2}{u_0^2}}}\sigma^1\tilde{y}^{0, l}_{\alpha s}
\\&-\alpha \frac{\left(k_x+2kl+\alpha k_l\right)\left(Q_{xx}^{1, 0}-Q_{yy}^{1,0}\right)}{\sqrt{2} u \sqrt{1 - \frac{u^2}{u_0^2}} k_l}\sigma^1\tilde{y}^{0, l}_{-\alpha \ 3-s}
\end{split}
 \ee
\be
\begin{split}
\mathcal{A}_{\alpha s}^{l\pm 1} =-&\frac{1}{4} Q_{zz}^{1,1}\partial_u\tilde{y}^{0, l}_{\alpha s}+ \frac{i\left[\frac{q_f}{\sqrt{3}} \left(\frac{u_0}{u}-1\right)Q_{tt}^{1,1}+2 q_f (u-u_0)A_{t}^{1,1}\right]}{4 u_0 \left(1 - \frac{u^2}{u_0^2}\right)}\sigma^3\tilde{y}^{0, l}_{\alpha s}-\alpha\frac{\pm k\left(k_x+2kl\right) Q_{zz}^{1,1}}{2\sqrt{2} k_{l} u}\sigma^1\tilde{y}^{0, l}_{\alpha s}
\\&-\frac{\alpha \left[\left(k_l^2-\left(k_x+2kl\right)^2\right) Q_{yy}^{1,1}+\left(k_x+2kl\right)\left(k_x\pm k+2kl\right) Q_{xx}^{1,1}\right]}{2 \sqrt{2} \ u \sqrt{1 - \frac{u^2}{u_0^2}}k_{l\mp 1}}
 \sigma^1\tilde{y}^{0, l}_{\alpha s}
\\&-\frac{\alpha\left(k_x+2kl+k_{l}\right) \left[\left(k_x\pm k +2kl)\right)Q_{xx}^{1,1}-(k_x+2kl)Q_{yy}^{1,1}\right]}{2 \sqrt{2} \ u \sqrt{1 - \frac{u^2}{u_0^2}} k_l}\sigma^1\tilde{y}^{0, l}_{-\alpha \ 3-s}
\\& \mp\frac{\alpha\left(k_x+2kl+k_{l}\right) k Q_{zz}^{1,1}}{2\sqrt{2} \ u \sqrt{1 - \frac{u^2}{u_0^2}} k_{l}}\sigma^1\tilde{y}^{0, l}_{-\alpha \ 3-s}~,
\end{split}
 \ee
After some straightforward algebra, we deduce
the second-order equation \be
\mathbf{L}_2 [ \tilde{y}_{\alpha s}^{1, l'} ] +X_{\alpha s}^{l'}=0~,
\ee
where $\mathbf{L}_2 [\tilde{y}]$ is defined in \eqref{secondordery2}, and
\be\label{hfunc}
X_{\alpha s}^{l'} =- \frac{\sqrt{2} k_l}{u\sqrt{1 - \frac{u^2}{u_0^2}}} \mathcal{A}_{\alpha s}^{l'} +i \frac{1+q_f A_t^0}{1- \frac{u^2}{u_0^2}} \mathcal{A}_{-\alpha \ 3-s}^{l'} + \frac{\alpha}{u\sqrt{1 - \frac{u^2}{u_0^2}}}  \partial_u \left( u\sqrt{1 - \frac{u^2}{u_0^2}} \mathcal{A}_{-\alpha \ 3-s}^{l'} \right)~.
\ee
The solution obeying the correct boundary condition at the horizon is
\be\label{eq58} \tilde{y}_{\alpha s}^{1,
	l'}(u)=-\tilde{y}_{\alpha s}^{0,l} (u)\int_u^\infty
du'\frac{\check{\tilde{y}}_{\alpha s}^{0,l}(u') X_{\alpha s}^{l'} (u')}{W(u')}+\check{\tilde{y}}_{\alpha s}^{0,l} (u) \int_u^\infty
du'\frac{\tilde{y}_{\alpha s}^{0,l} (u')X_{\alpha s}^{l'} (u')}{W(u')}~,
\ee
where $\tilde{y}_{\alpha s}^{0,l}$ is the zeroth-order solution satisfying ingoing boundary conditions at the horizon (Eqs.\ \eqref{yInOut} and \eqref{zInOut}),  $\check{\tilde{y}}_{\alpha s}^{0,l}$ is a linearly independent zeroth-order solution satisfying outgoing boundary conditions at the horizon, and $W$ is their Wronskian.

This first-order solution ought to be matched with the solution in the far region. Its asymptotic expression in the overlap region as a function of the frequency includes terms which behave as $\omega^0$, $\omega^{2\nu_{k_{l'}}}$, and $\omega^{\nu_{k_{l'}} \pm \nu_{k_l}}$, respectively. The potentially divergent terms proportional to $\omega^{\nu_{k_{l'}} - \nu_{k_l}}$ do not contribute to the Green function, because they can be absorbed into the overall (physically irrelevant) normalization of the solutions to the Dirac equation.

Care must be exercised in the important special case of degeneracy ($\nu_{k_{l'}} = \nu_{k_l}$), which is relevant to the calculation of the gap. The solution in this case can be found by carefully taking the limit $\nu_{k_{l'}} \to \nu_{k_l}$. We obtain a solution whose asymptotic expression in the overlap region contains terms which behave as $\omega^0$, $\omega^{2\nu_{k_{l}}}$, and $\omega^{2 \nu_{k_l}} \ln \omega$, respectively, all of which are well-behaved in the limit $\omega \to 0$.

 We also calculated the first-order solution to the Dirac equation numerically and found good agreement with the analytic expressions. However, a calculation of the retarded Green function revealed no gap below the critical temperature. To see the pseudogap, we need to include second-order effects which we proceed to calculate next.

\subsection{Second Order}
\label{DiracSecondAna}

At second order, i.e., at $\mathcal{O} (\xi^4)$, below the critical temperature, five modes are excited, with $l' = l,l\pm 1, l\pm 2$. For an analytic solution, we need to analyze the near-horizon region, as in the first-order case. After performing the rotation described in Appendix \ref{spinorrota} at second order,
\be \left( \begin{array}{c} \tilde{\psi}_{+s}^{2,l} \\ \tilde{\psi}_{-s}^{2,l} \end{array} \right) = \left( \begin{array}{cc} \cos\frac{\theta}{2} & - \sin\frac{\theta}{2} \\ \sin\frac{\theta}{2} & \cos\frac{\theta}{2} \end{array} \right) \left( \begin{array}{c} {\psi}_{+s}^{2,l} \\ {\psi}_{-s}^{2,l} \end{array} \right) \ee
where $\tan\theta = \frac{k_y}{k_x+2kl}$, and defining
\be\label{phitilde2}
\tilde{\psi}_{\alpha s}^{2,l} =\frac{1}{\sqrt{2}}\left( 1-\frac{u^2}{u_0^2} \right)^{-1/4} (1+i \sigma^1)\tilde{y}_{\alpha s}^{2,l} (u)~.
\ee
the second-order massless Dirac equation near the horizon becomes
\be\label{eq2Dirac}
\partial_u \tilde{y}_{\alpha s}^{2,
	l'}	-\frac{i}{ 1 - \frac{u^2}{u_0^2}} \sigma^3\left( 1 + q_f
A_{t}^{0}\right)\tilde{y}_{\alpha s}^{2, l'}-\frac{ \alpha \sqrt{2}k_l}{\mu_0u\sqrt{ 1 - \frac{u^2}{u_0^2}}
} \sigma^1 \tilde{y}_{\alpha s}^{2, l'} + \mathcal{B}_{\alpha s}^{l'}=0~,
\ee
where
\be
\begin{split}
 \mathcal{B}_{\alpha s}^{l}&= \frac{3}{16} \left(2 {Q_{zz}^{1,0}}^2+{Q_{zz}^{1,1}}^2\right)\partial_u \tilde{y}^{0, l}_{\alpha s}-\frac{1}{2} Q_{zz}^{1,0} \partial_u \tilde{y}^{1, l}_{\alpha s}-\frac{1}{2}Q_{zz}^{1,0} \partial_u \tilde{y}^{1, l}_{\alpha s}-\frac{1}{4}Q_{zz}^{1,1} \partial_u \tilde{y}^{1, l-1}_{\alpha s}-\frac{1}{4}Q_{zz}^{1,1} \partial_u \tilde{y}^{1, l+1}_{\alpha s}
\\&+\frac{i  Q_{tt}^{1,0} \left[q_f (u-u_0) \left(\sqrt{3}   Q_{tt}^{1,0}-4 u A_t^{1,0}\right)-3 u u_0 Q_{tt}^{1,0}\right]}{8 u u_0 \left(1 - \frac{u^2}{u_0^2}\right)}\sigma^3 \tilde{y}_{\alpha s}^{0, l} 
\\&+\frac{i  Q_{tt}^{1,1} \left[Q_{tt}^{1,1} \left(\sqrt{3} q_f (u-u_0)-3 u u_0\right)+4 q_f u (u_0-u) A_{t}^{1,1}\right]}{16 u u_0 \left(1 - \frac{u^2}{u_0^2}\right)}\sigma^3 \tilde{y}_{\alpha s}^{0, l} 
\\&+\sum_{l'=l\pm1}\frac{i  Q_{tt}^{1,1} \left(\sqrt{3} q_f (u_0-u)+3 u u_0\right)+6 q_f u (u-u_0) A_{t}^{1,1}}{12 u u_0 }\sigma^3 \tilde{y}_{\alpha s}^{1, l'} 
\\&+\frac{i  Q_{tt}^{1,0} \left(\sqrt{3} q_f (u_0-u)+3 u u_0\right)+6 q_f u (u-u_0) A_{t}^{1,0}}{12 u u_0 }\sigma^3 \tilde{y}_{\alpha s}^{1, l} 
\\&-\alpha \sum_{l'=l\pm1}\frac{Q_{xx}^{1,1} (2 k l'+k_x) (k (2 l+1)+k_x)-Q_{yy}^{1,1}\left((2 k l'+k_x)^2-k_{l'}^2\right)-k  Q_{zz}^{1,1} (2 k l'+k_x)}{2 \sqrt{2\left(1 - \frac{u^2}{u_0^2}\right)} \mu_0 u k_{l'}}\sigma^1 \tilde{y}_{\alpha s}^{1, l'} 
\\&+\frac{3 \alpha  \left(-\left(2 {Q_{yy}^{1,0}}^2+{Q_{yy}^{1,1}}^2\right) \left((2k l+k_1)^2-k_l^2\right)+\left(2 {Q_{xx}^{1,0}}^2 +{Q_{xx}^{1,1}}^2 \right)(2k l+k_1)^2\right)}{8 \sqrt{2\left(1 - \frac{u^2}{u_0^2}\right)} \mu_0 u k_l} \sigma^1 \tilde{y}_{\alpha s}^{0, l} 
\\&-\alpha \frac{Q_{xx}^{1,0} (2k l+k_1)^2-Q_{yy}^{1,0} \left((2k l+k_1)^2-k_l^2\right)}{\sqrt{2 \left(1 - \frac{u^2}{u_0^2}\right)} \mu_0 u k_l}\sigma^1 \tilde{y}_{\alpha s}^{1, l} 
\\&+\alpha\frac{3 (2k l+k_1) (2k l+k_1+k_l) \left(2 {Q_{xx}^{1,0}}^2+{Q_{xx}^{1,1}}^2-2 {Q_{yy}^{1,0}}^2-{Q_{yy}^{1,1}}^2\right)}{8 \sqrt{2 \left(1 - \frac{u^2}{u_0^2}\right)} \mu_0 u k_l}\sigma^1 \tilde{y}_{-\alpha 3-s}^{0, l} 
\\&-\alpha\frac{(2k l+k_1) (Q_{xx}^{1,0}-Q_{yy}^{1,0}) (2k l+k_x+k_l)}{\sqrt{2\left(1 - \frac{u^2}{u_0^2}\right)} \mu_0 u k_l}\sigma^1 \tilde{y}_{-\alpha 3-s}^{1, l} 
\\&-\alpha\sum_{l'=l\pm1}\frac{((2k l'+k_1\mp k) Q_{xx}^{1,1}-(2k l'+k_1) Q_{yy}^{1,1}\pm kQ_{zz}^{1,1}) (2k l'+k_x+k_l')}{2\sqrt{2\left(1 - \frac{u^2}{u_0^2}\right)} \mu_0 u k_l'}\sigma^1 \tilde{y}_{-\alpha 3-s}^{1, l'} 
\end{split}
\ee
\be
\begin{split}
\mathcal{B}_{\alpha s}^{l\pm 1}&=\frac{3}{8} Q_{zz}^{1,0} Q_{zz}^{1,1}\partial_u \tilde{y}^{0, l}_{\alpha s}-\frac{Q_{zz}^{1,1}}{4}\partial_u \tilde{y}^{1, l}_{\alpha s}-\frac{Q_{zz}^{1,1}}{2}\partial_u \tilde{y}^{1, l+1}_{\alpha s}
\\&+\frac{i \left(Q_{tt}^{1,0} \left( Q_{tt}^{1,1} \left(\sqrt{3} q_f (u-u_0)-3 u u_0\right)+2 q_f u (u_0-u) A_t^{1,1}\right)+2 q_f u Q_{tt}^{1,1} (u_0-u) A_t^{1,0}\right)}{8 u u_0 \left(1 - \frac{u^2}{u_0^2}\right)}\sigma^3 \tilde{y}^{0, l}_{\alpha s}
\\&+\frac{i  \left(Q_{tt}^{1,1} \left(\sqrt{3} q_f (u_0-u)+3 u u_0\right)+6 q_f u (u-u_0) A_{t}^{1,1}\right)}{12 u u_0 \left(1 - \frac{u^2}{u_0^2}\right)} \sigma^3\tilde{y}^{1, l}_{\alpha s}
\\&+\frac{i  \left(Q_{tt}^{1,0} \left(\sqrt{3} q_f (u_0-u)+3 u u_0\right)+6 q_f u (u-u_0) A_{t}^{1,0}\right)}{6 u u_0 \left(1 - \frac{u^2}{u_0^2}\right)} \sigma^3\tilde{y}^{1, l+1}_{\alpha s}
\\&+\alpha \frac{Q_{xx}^{1,0} (2kl+k_x) (3 Q_{xx}^{1,1} (\mp k+2 k l+k_x)+k Q_{zz}^{1,1})}{4 \sqrt{2\left(1 - \frac{u^2}{u_0^2}\right)} \mu_0 u k_{l}} \sigma^1 \tilde{y}^{0, l}_{\alpha s}
\\&-\alpha \frac{ 3 Q_{yy}^{1,0} Q_{yy}^{1,1} \left((2kl+k_x)^2-k_{l}^2\right)-2 k Q_{zz}^{1,0} Q_{zz}^{1,1} (2kl+k_x)}{4 \sqrt{2\left(1 - \frac{u^2}{u_0^2}\right)} \mu_0 u k_{l}}\sigma^1 \tilde{y}^{0, l}_{\alpha s}
\\&+\alpha\frac{Q_{xx}^{1,1} (-(2 kl+k_x)) (\mp k+2kl+k_x)-k Q_{zz}^{1,1} (2kl +k_x)+Q_{yy}^{11} \left((2kl+k_x)^2-k_{l}^2\right)}{2 \sqrt{2\left(1 - \frac{u^2}{u_0^2}\right)} \mu_0 u k_{l\pm 1}}\sigma^1 \tilde{y}^{1, l}_{\alpha s}
\\&-\alpha \frac{ (k_x+2(k \pm1))^2Q_{xx}^{1,0}+\left((2k(l\pm 1)+k_1)^2-k_{l\pm1}^2\right)Q_{yy}^{1,0}}{ \sqrt{2\left(1 - \frac{u^2}{u_0^2}\right)} \mu_0 u k_{l\pm 1}}\sigma^1 \tilde{y}^{1, l\pm1}_{\alpha s}
\\&+\alpha\frac{Q_{xx}^{1,0} ((2 k l+k_x+k_{l}) (3 Q_{xx}^{1,1}(2 k l\pm k+k_1)+k Q_{zz}^{1,1})}{4 \sqrt{2\left(1 - \frac{u^2}{u_0^2}\right)} \mu_0 u k_{l\pm1}}\sigma^1 \tilde{y}^{0, l}_{-\alpha 3-s}
\\&-\alpha\frac{((2 k l+k_1+k_l) ( \pm 3 Q_{yy}^{1,0} Q_{yy}^{1,1} (2 k l+k_x)\mp 2 k Q_{zz}^{1,0} Q_{zz}^{1,1})}{4 \sqrt{2\left(1 - \frac{u^2}{u_0^2}\right)} \mu_0 u k_{l\pm 1}}\sigma^1 \tilde{y}^{0, l}_{-\alpha 3-s}
\\&-\alpha\frac{ (2 k l+k_x+k_l) ((2 k l+k+k_x)Q_{xx}^{1,1} - (k_x+2kl)Q_{yy}^{1,1}-kQ_{zz}^{1,1})}{2\sqrt{2 \left(1 - \frac{u^2}{u_0^2}\right)} \mu_{0} u k_{l\pm 1}}\sigma^1 \tilde{y}^{1, l}_{-\alpha 3-s}
\\&-\alpha\frac{ (2 k (l\pm 1)+k_x) (2 k (l\pm1)+k_{l\pm1}+k_x)(Q_{xx}^{1,0} -Q_{yy}^{1,0})}{\sqrt{2 \left(1 - \frac{u^2}{u_0^2}\right)} \mu_{0} u k_{l\pm 1}}\sigma^1 \tilde{y}^{1, l\pm1}_{-\alpha 3-s}
\end{split}
\ee
\be 
\begin{split}
\mathcal{B}_{\alpha s}^{l\pm 2}&=-\frac{1}{4}Q_{zz}^{1,1}\partial_u \tilde{y}^{1, l\pm 1}_{\alpha s}+\frac{i  \left(Q_{tt}^{1,1} \left(\sqrt{3} q_f (u_0-u)+3 u u_0\right)+6 q_f u (u-u_0) A_t^{1,1}\right)}{12 u u_0\left(1 - \frac{u^2}{u_0^2}\right)}\tilde{y}^{1, l\pm 1}_{\alpha s}
\\&+\alpha\frac{Q_{yy}^{1,1} \left((2 k (l\pm1)+k_x)^2-k_{l\pm1}^2\right)+Q_{xx}^{1,1} (-(2 k (l\pm1)+k_x)) (k (2 l-3)+k_x)}{2 \sqrt{2\left(1 - \frac{u^2}{u_0^2}\right)} \mu_0 u k_{l\pm1}}\sigma^1 \tilde{y}^{1, l\pm1}_{\alpha s}
\\&-\alpha\frac{k Q_{zz}^{1,1} (2 k (l\pm 1)+k_x)}{2 \sqrt{2\left(1 - \frac{u^2}{u_0^2}\right)} \mu_0 u k_{l\pm1}}\sigma^1 \tilde{y}^{1, l\pm1}_{\alpha s}
\\&-\frac{  (2 k (l\pm 1)+k_x+k_{l\pm1}) (Q_{xx}^{1,1} (k (2 l-3)+k_x)-Q_{yy}^{1,1} (2 k (l\pm 1)+k_x)+k Q_{zz}^{1,1})}{2 \sqrt{2\left(1 - \frac{u^2}{u_0^2}\right)} \mu_0 u k_{l\pm1}}\sigma^1 \tilde{y}^{1, l\pm1}_{-\alpha\ 3-s}
\end{split}
\ee
After some algebra, we arrive at the second order equation
\be
\mathbf{L}_2 [ \tilde{y}_{\alpha s}^{2, l'} ] +Y_{\alpha s}^{l'}=0~,
\ee
where $\mathbf{L}_2 [\tilde{y}]$ is defined in \eqref{secondordery2}, and
\be Y_{\alpha s}^{l'}= - \frac{\sqrt{2} k_l}{u\sqrt{1 - \frac{u^2}{u_0^2}}} \mathcal{B}_{\alpha s}^{l'} +i \frac{1+q_f A_t^0}{1- \frac{u^2}{u_0^2}} \mathcal{B}_{-\alpha \ 3-s}^{l'} + \frac{\alpha}{u\sqrt{1 - \frac{u^2}{u_0^2}}}  \partial_u \left( u\sqrt{1 - \frac{u^2}{u_0^2}} \mathcal{B}_{-\alpha \ 3-s}^{l'} \right)~. \ee
The solution obeying the correct boundary condition at the horizon is (\emph{cf.}\ with its first-order counterpart \eqref{eq58})
\be \tilde{y}_{\alpha s}^{2,
	l'}(u)=-\tilde{y}_{\alpha s}^{0,l} (u)\int_u^\infty
du'\frac{\check{\tilde{y}}_{\alpha s}^{0,l}(u') Y_{\alpha s}^{l'} (u')}{W(u')}+\check{\tilde{y}}_{\alpha s}^{0,l} (u) \int_u^\infty
du'\frac{\tilde{y}_{\alpha s}^{0,l} (u')Y_{\alpha s}^{l'} (u')}{W(u')}~.
\ee
As before, this second-order solution ought to be matched with the solution in the far region. Its asymptotic expression in the overlap region as a function of the frequency includes terms which behave as $\omega^0$, $\omega^{2\nu_{k_{l'}}}$, and $\omega^{\nu_{k_{l'}} \pm \nu_{k_l}}$, respectively. The potentially divergent terms, $\omega^{\nu_{k_{l'}} - \nu_{k_l}}$, may be absorbed into the overall normalization of the solutions, as in first order.

In the special case of degeneracy ($\nu_{k_{l'}} = \nu_{k_l}$), which contributes to the gap at second order, the solution can be found by carefully taking the limit $\nu_{k_{l'}} \to \nu_{k_l}$. We obtain a solution whose asymptotic expression in the overlap region contains terms which behave as $\omega^0$, $\ln\omega$, $\omega^{2\nu_{k_{l}}}$, $\omega^{2 \nu_{k_l}} \ln \omega$, and $\omega^{2 \nu_{k_l}} (\ln \omega)^2$, respectively. The terms proportional to $\ln\omega$ diverge in the limit $\omega \to 0$, however an explicit calculation shows that they do not contribute to the pole at the Fermi surface ($k_l = k_F$).

We calculated the second-order solution numerically and found good agreement with the analytic expressions. A calculation of the retarded Green function revealed no gap in the general case.  However, a pseudogap emerged in the degenerate case, as we discuss next.

\subsection{Generation of a gap in the Fermi surface}
\label{bandgapAna}

\begin{figure}[t]
\includegraphics[width=.5\textwidth]{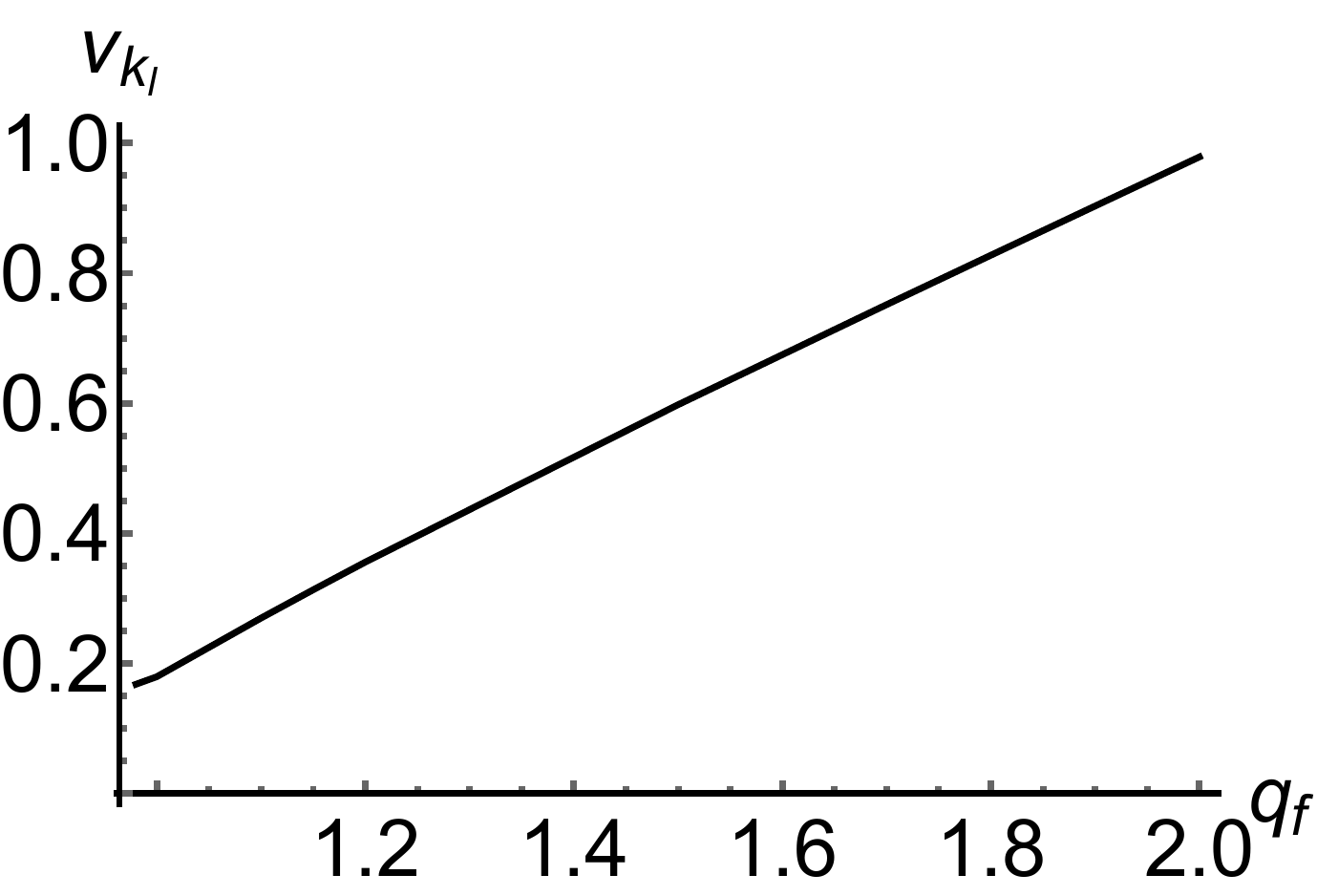}
\caption{Plot of scaling parameter, $\nu_{k_l}$, as a function of
fermion charge, $q_f$, at $k_l=k_F$ for parameters
$\frac{\eta}{\mu_0^2}=0.25$, $\frac{\eta'}{\mu_0^4}=0.005$, $q=0$,
$\Delta=1$, $\mu_0=2.2507$, $\frac{T}{\mu_0}=0.0613$, and tiny
frequecy $\omega=0.0001$  respectively. } \label{nu_qf}
\end{figure}

\begin{figure}[t]
\includegraphics[width=.45\textwidth]{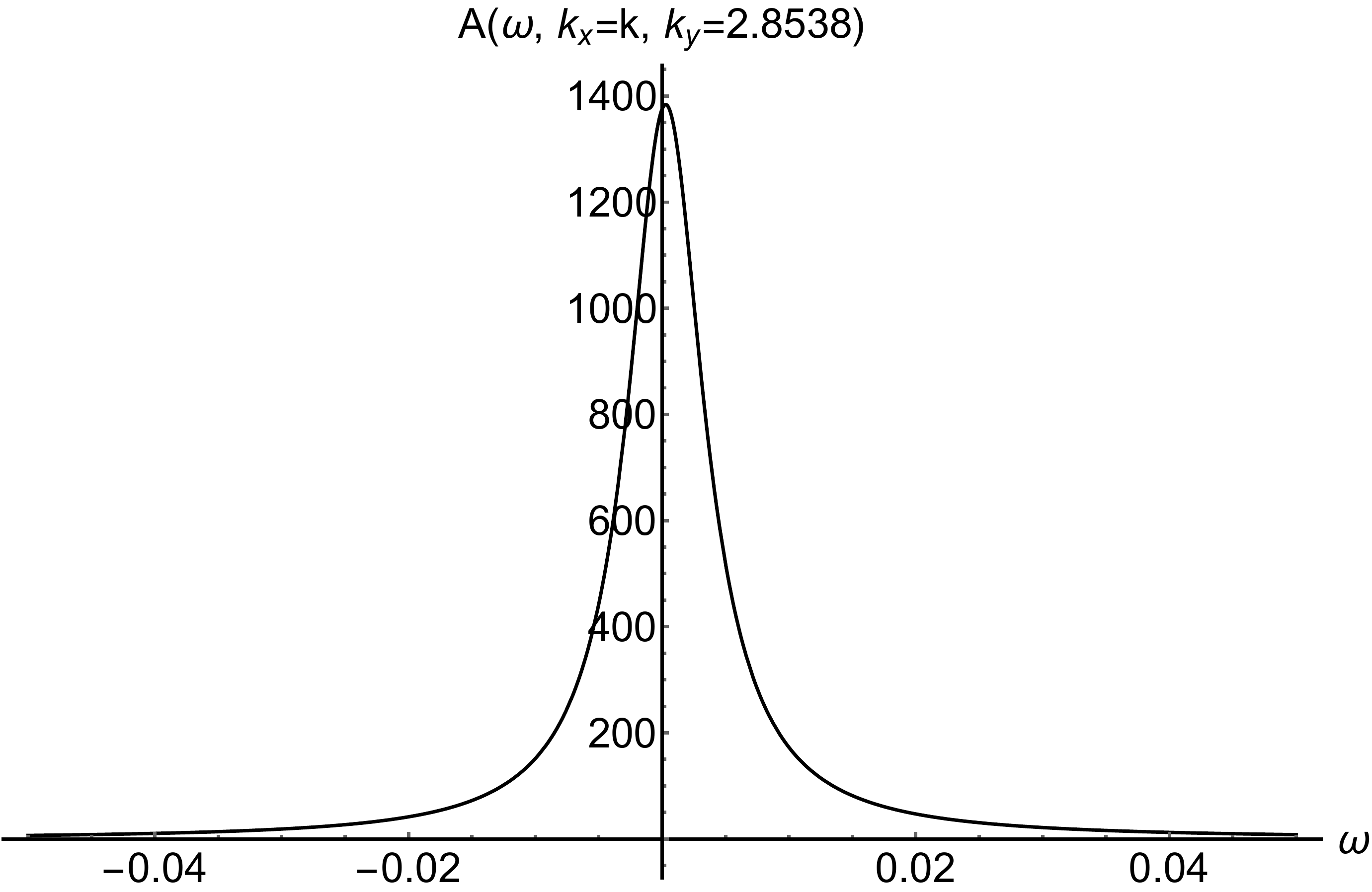}
\includegraphics[width=.45\textwidth]{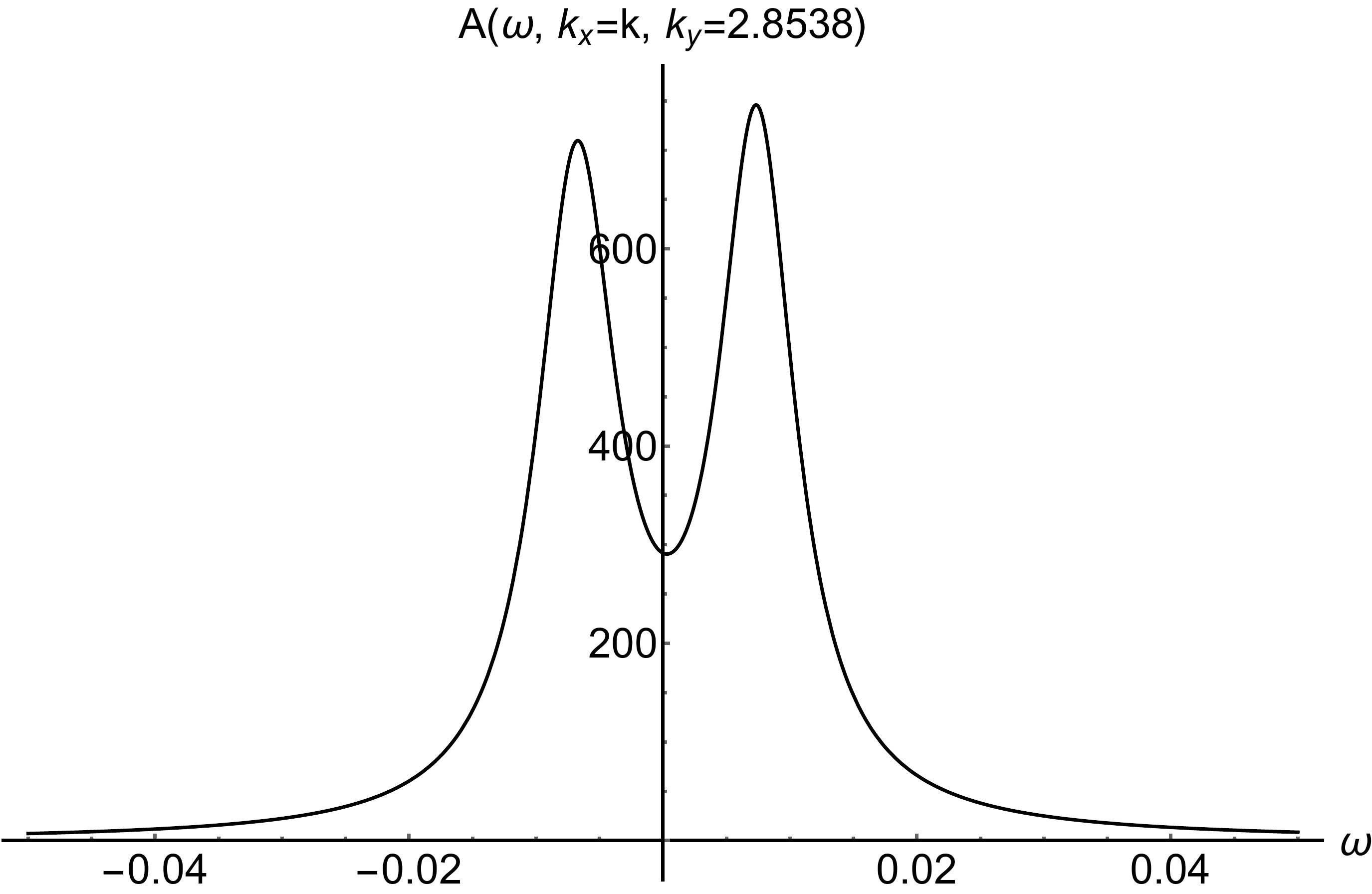} \\
\includegraphics[width=.45\textwidth]{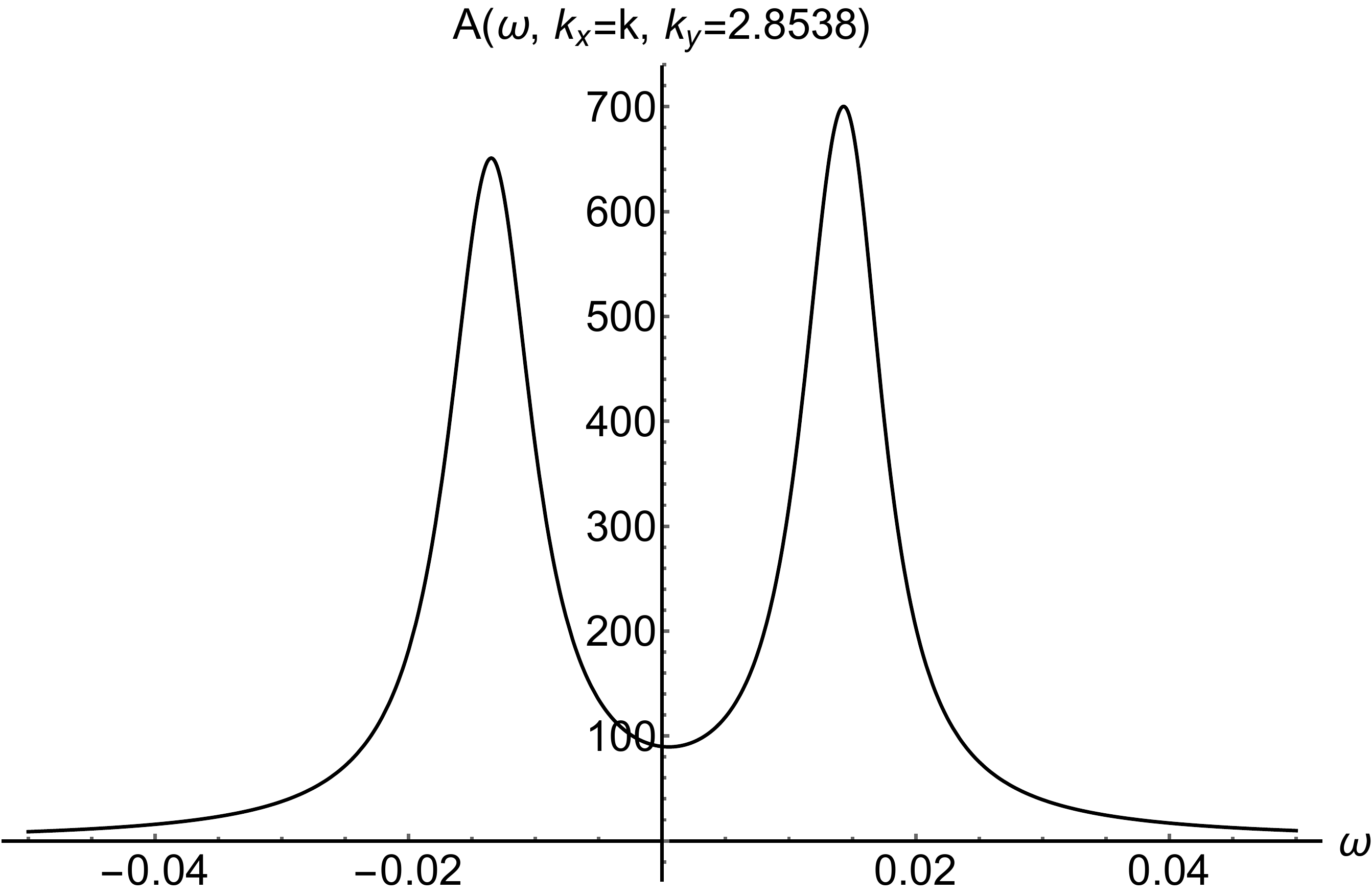}
\includegraphics[width=.45\textwidth]{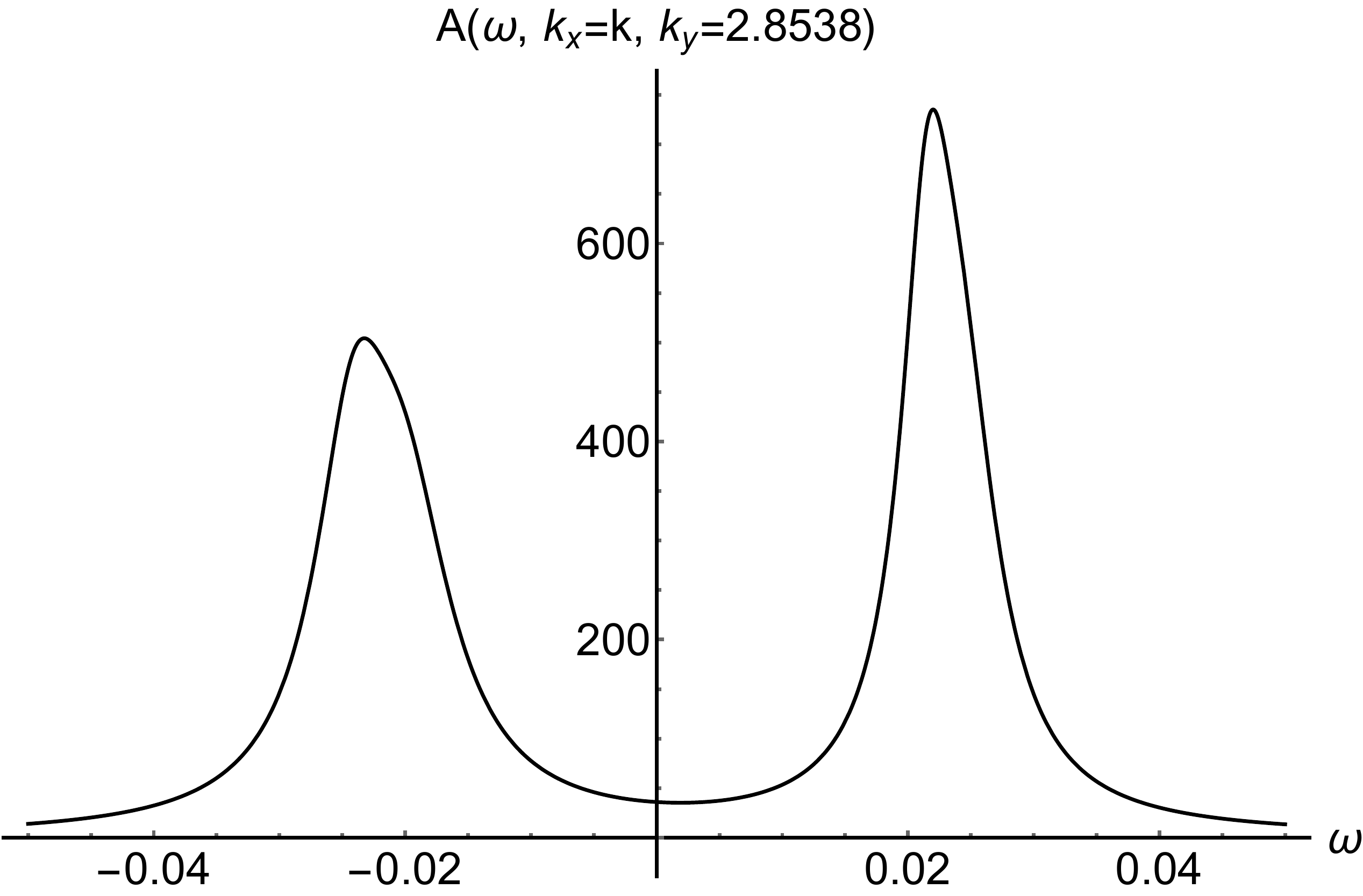}
\caption{Plot of spectral function $A(\omega, k_x=k, k_y=2.8538)$ vs. $\omega$ for parameters
$\frac{\eta}{\mu_0^2}=0.25$, $\frac{\eta'}{\mu_0^4}=0.005$, $q=0$,
$\Delta=1$, $q_f=2$, $\mu_0=2.2507$, $\frac{T}{\mu_0}=0.0613$ and for $\xi=0.0, 0.05, 0.07, 0.1$ respectively. } \label{gapomega}
\end{figure}

As already discussed, the retarded Green function does not develop any pole at the Fermi surface using the general first and second order solutions of the Dirac equation. However, in the degenerate case, $\nu_{k_{l'}} = \nu_{k_l}$ of the second order solution,
a gap is generated in the Fermi surface corresponding to a pole in the  Green function  where
$k_l=k_F$.

 Near the Fermi surface, we can write the
Green function as in \cite{LSSZ:2012},
\be \left( G_{R} \right)_{ \alpha l, \alpha l}=\frac{A_{\alpha l, \alpha l}^{(0)}
B_{\alpha l, \alpha l}^{(0)}}{(A_{\alpha l, \alpha l}^{(0)} +\xi^4
A_{\alpha l, \alpha l}^{(2)} )^2-\xi^4 A_{\alpha l, \alpha
l-1}^{(1)} A_{\alpha l-1, \alpha l}^{(1)} +\mathcal{O}(\xi^6)} \label{Greensfunc}~,
\ee
where $A^{(n)}_{\alpha l, \alpha l}$ and $B^{(n)}_{\alpha l, \alpha l}$ are the results of matching at $n$-th order in the boundary behavior \eqref{boundarydirac}.
To find a numerical solution to the Green function \eqref{Greensfunc}, we solve the second order
Dirac equation with ingoing boundary conditions at the horizon and plot the spectral function as a function of $\omega$ and $k_y$ respectively. The results of our calculations are depicted in Fig. \ref{gapomega} - Fig. \ref{nflgapomega}  and discussed in details in the following.

Near the Fermi surface, the retarded Green functions take the general form
\be G_R=\frac{Z}{\omega - v_F(k-k_F)+\Sigma(\omega, k)}~, \ee where
$\Sigma$ is the self energy for fermionic excitations near the Fermi
surface and $Z$ is the residue of the pole and
quasiparticle weight. Therefore, we can write
\be A_{\alpha l, \alpha
l}^{(0)} +\xi^4 A_{\alpha l, \alpha l}^{(2)}
=\omega-v_F(k_l-k_F)+i(c_1-ic_2)\omega^{2\nu_{k_l}}\ee where $v_F,
c_1, c_2$ are real constants determined from the boundary
data.

We can distinguish three different cases for the  retarded Green
function  depending on the value of $\nu_{k_l}$ \cite{Faulkner:2009wj}. The $\nu_{k_l}$ values are plotted as a function of $q_f$  in Fig. \ref{nu_qf}.

\begin{itemize}

\item For $\nu_{k_l} > \frac{1}{2}$, the system has fermionic
quasiparticles and the effective theory  is a Fermi liquid. In
this case the imaginary part of the self-energy is proportional to $ \omega^2$. Also, the spectral function has a
 Lorentzian distribution centered around $\omega=0$. The dominant linear term leads the
dispersion and we obtain 
\be G_{R \alpha l,\alpha l}^{-1} \sim
(\omega-v_F(k_l-k_F))^2-\Delta^2+i c_1
(\omega-v_F(k_l-k_F))\omega^{2 \nu_{k_l}}~. \ee 
Near the Fermi
surface and with small $\omega$, there are two peaks in the spectral
function, $A(\omega, k_x, k_y)=\Im[G_{R1l, 1l}+G_{R 2l, 2l}]$. The peaks are found at $\omega=v_F(k_l-k_F)\pm \Delta$ as seen in
Fig. {\ref{gapomega}}, and the pseudogap $\Delta$ is first order
in $\xi^2$, given by \be \Delta^2=\xi^4 A_{\alpha l, \alpha
l-1}^{(1)} A_{\alpha l-1, \alpha l}^{(1)}~. \ee  The width is controlled by  second-order terms. Fig.\
\ref{gapomega} also shows that the size of the gap is on the
order of $\xi^2$, as $\omega \sim \Delta \sim \xi^2$. 
Also apparent is the pseudogap behavior as the spectral function remains non-zero over all energies.
The value $\xi=0$
corresponds to temperatures above critical temperature. As we increase
$\xi$, the magnitude of the gap increases but at large enough $\xi$, the perturbation breaks down. The widths are on the order of $\xi^{2\nu_{k_l}}$, therefore, they  appear as sharp peaks. For the parameters
used in plotting Fig.\ \ref{gapomega}, the system is in the Fermi liquid state with $\nu_{k_l}=0.9787$.

\begin{figure}[t]
\begin{center}
\includegraphics[width=.45\textwidth]{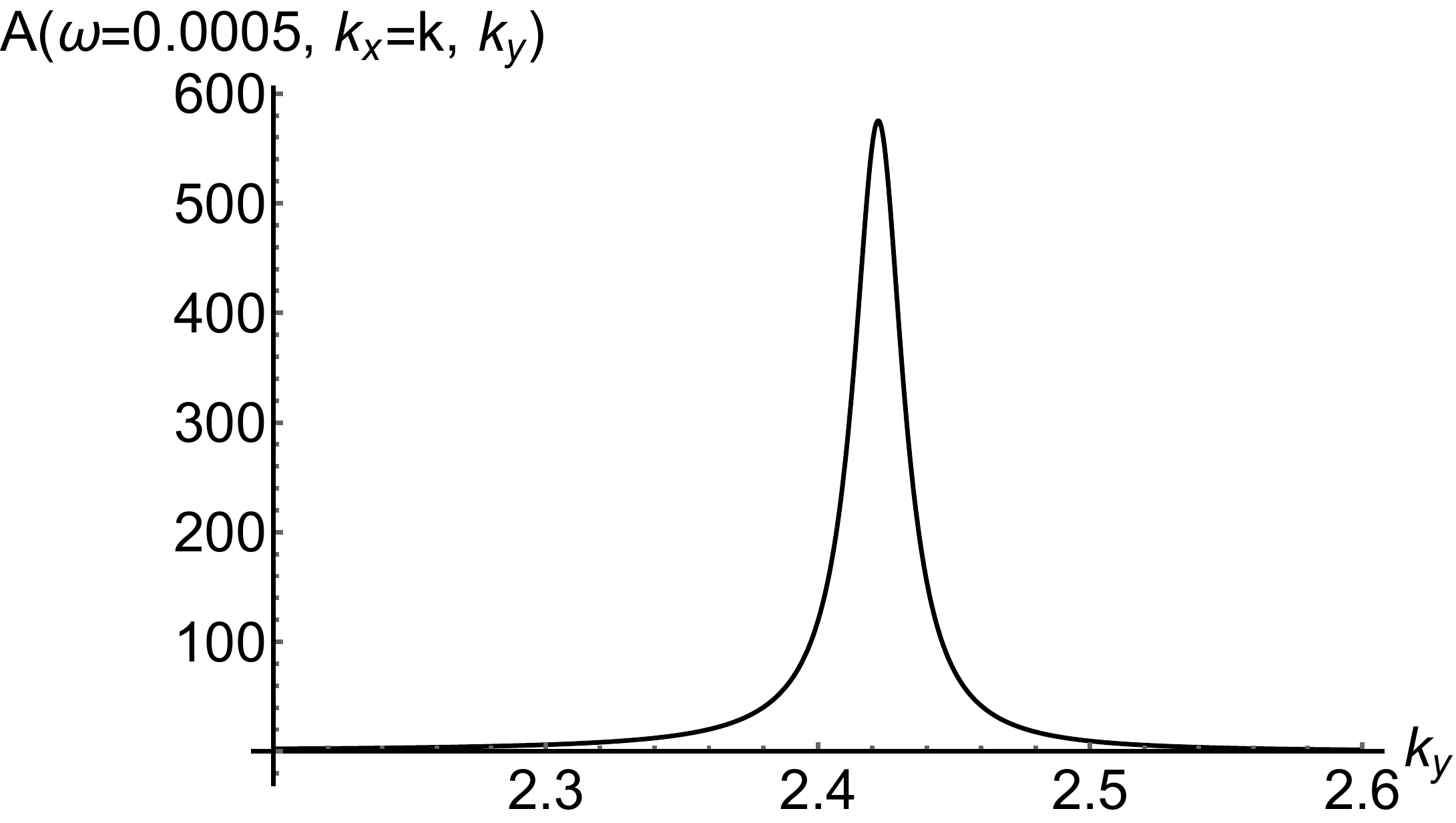}
\includegraphics[width=.45\textwidth]{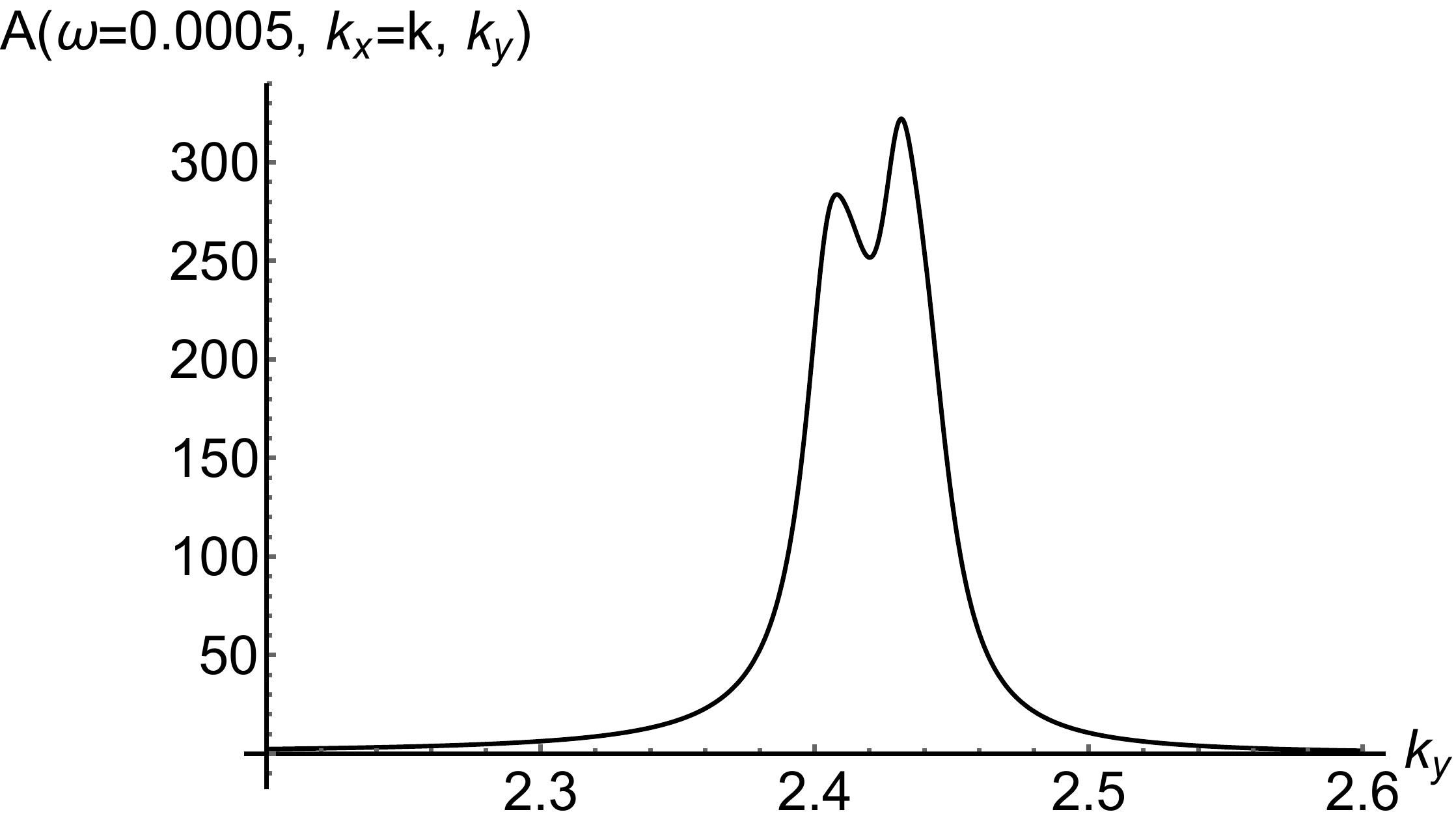} \\
\includegraphics[width=.45\textwidth]{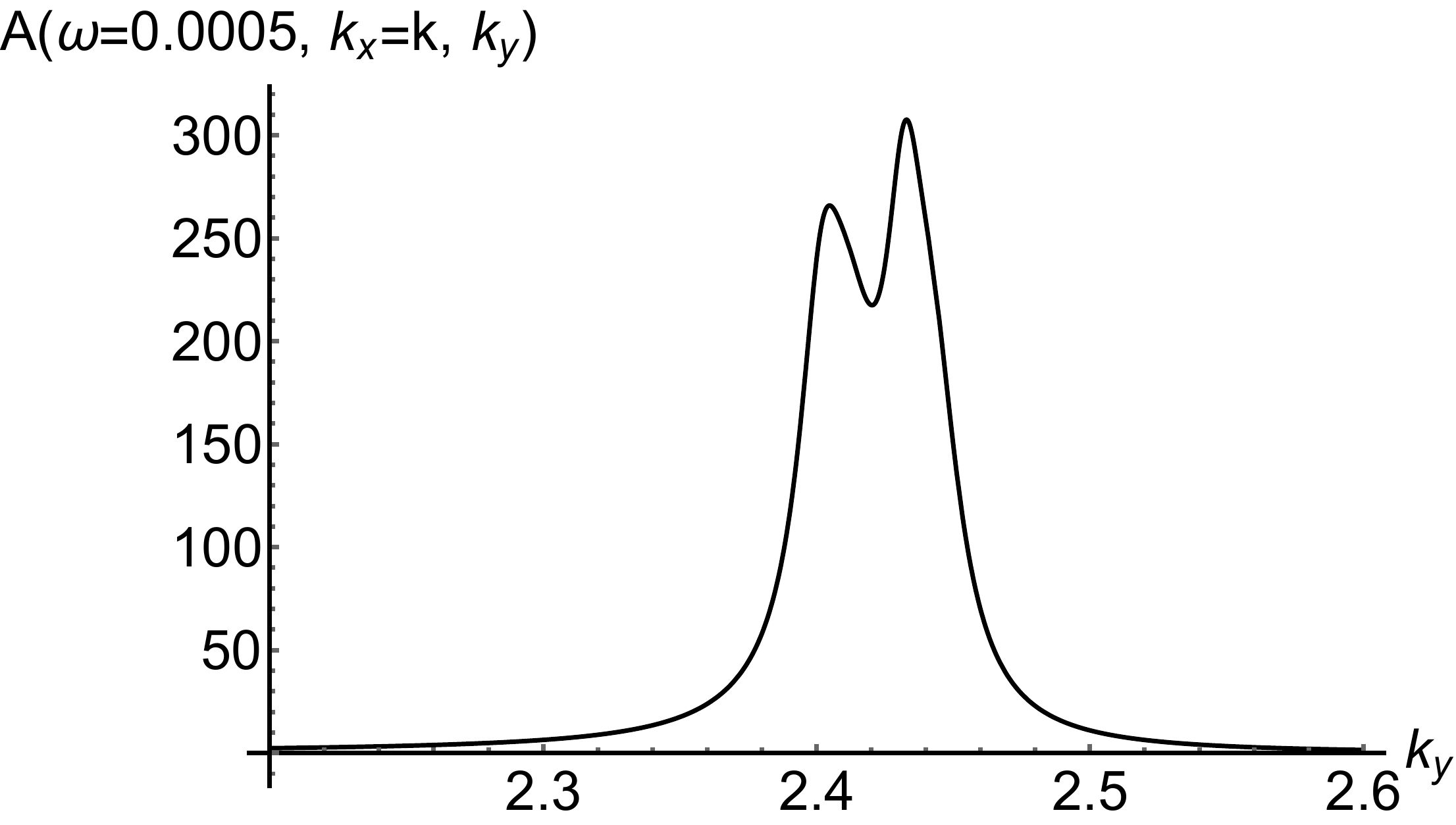}
\includegraphics[width=.49\textwidth]{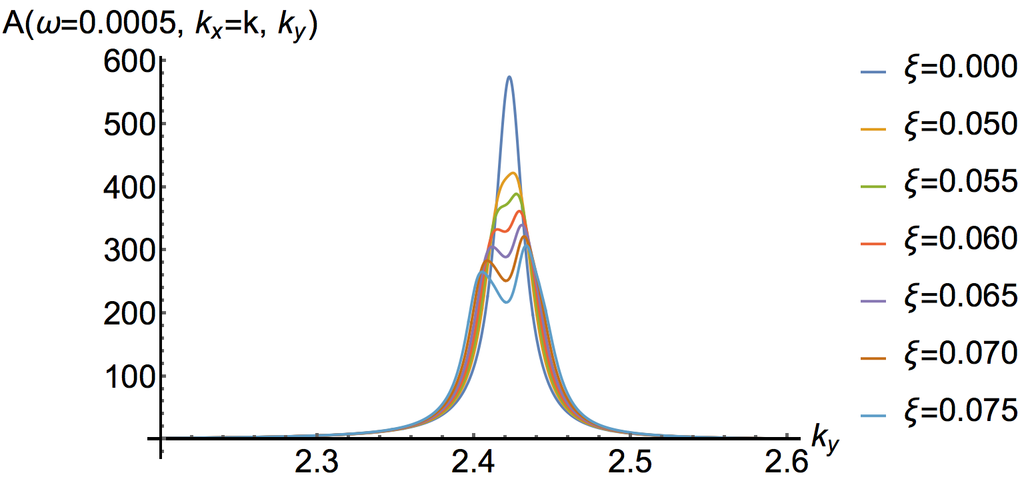}
\caption{Plot of spectral function $A(\omega=0.0005, k_x=k, k_y)$ vs. $k_y$ for parameters
$\frac{\eta}{\mu_0^2}=0.25$, $\frac{\eta'}{\mu_0^4}=0.005$, $q=0$,
$\Delta=1$, $q_f=1.8$, $\mu_0=2.2507$, $\frac{T}{\mu_0}=0.0613$ and for $\xi=0.0, 0.07, 0.075$ respectively. } \label{gapky}
\end{center}
\end{figure}

We  also observe the presence of the pseudogap in the $A(\omega, k_x,
k_y)$ \emph{vs}.\ $k_y$ graph in Fig.\ \ref{gapky} for different values of $\xi$.
 As seen from the graphs, the perturbation breaks down near $\xi \sim 0.08$.

\item For $\nu_{k_l} < \frac{1}{2}$ in the
non-Fermi liquid  case, the non-analytic term dominates non-linear
dispersion and we can write $A_{\alpha l, \alpha l}^{(0)} +\xi^4
A_{\alpha l, \alpha l}^{(2)} =c_2 \omega^{2
\nu_{k_l}}-v_F(k_l-k_F)+ic_1-\omega^{2\nu_{k_l}}$.  This produces the
 Green function 
 \be G_{R \alpha l,\alpha l}^{-1} \sim
(\omega^{2\nu_{k_l}}-v_F(k_l-k_F))^2-\Delta^2+i
c_1 (\omega^{2\nu_{k_l}}-v_F(k_l-k_F))\omega^{2 \nu_{k_l}}~, \ee
and the two peaks are at $\omega^{2\nu_{k_l}}=v_F(k_l-k_F)\pm
\Delta$. This qualitatively differs from that of the Fermi liquid case.   For the non-Fermi liquid case, the width of
the non-linear dispersion and gap are on the order of
$\xi^{1/2\nu_{k_l}}$.

\begin{figure}[t]
\includegraphics[width=.45\textwidth]{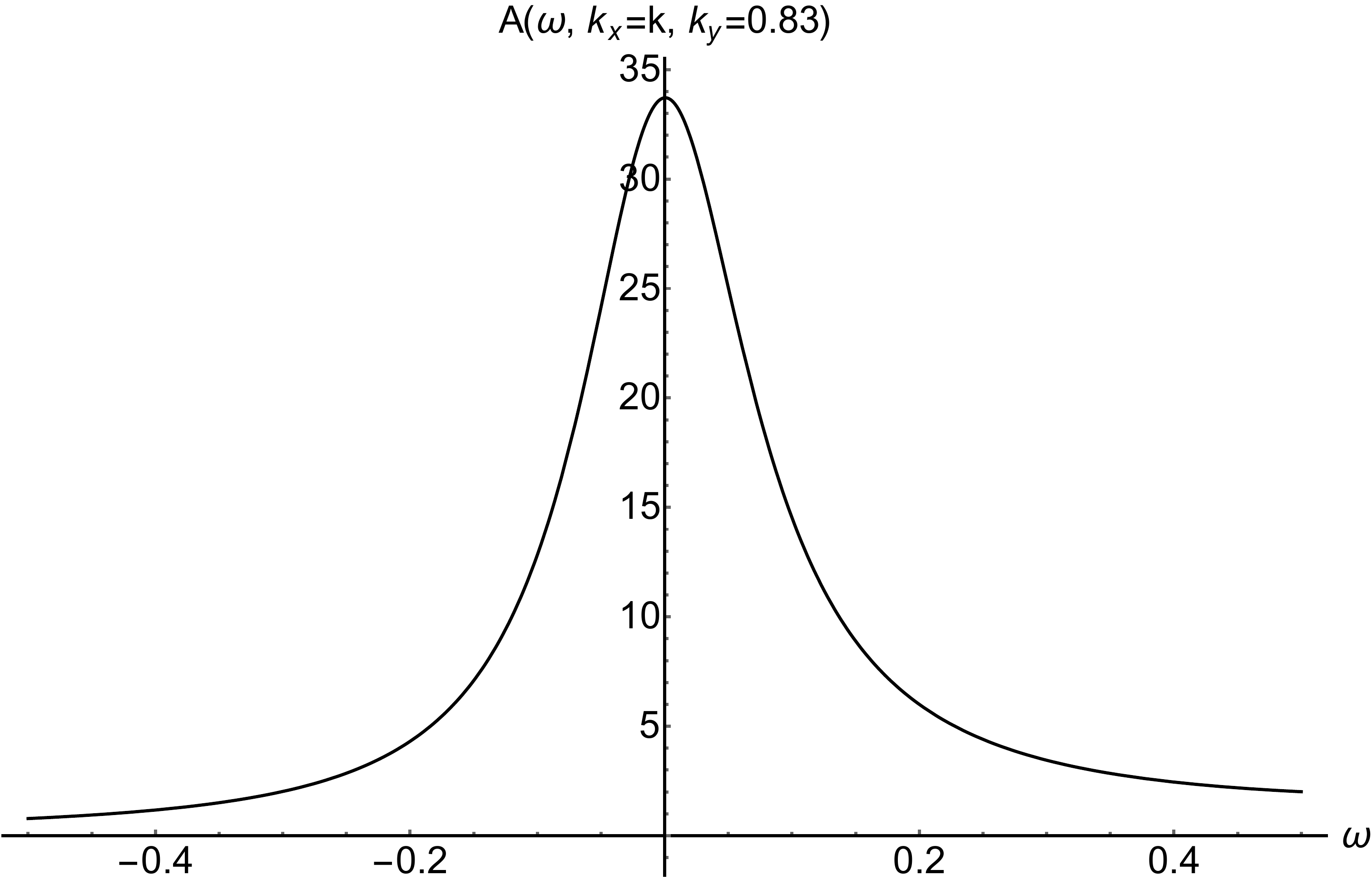}
\includegraphics[width=.45\textwidth]{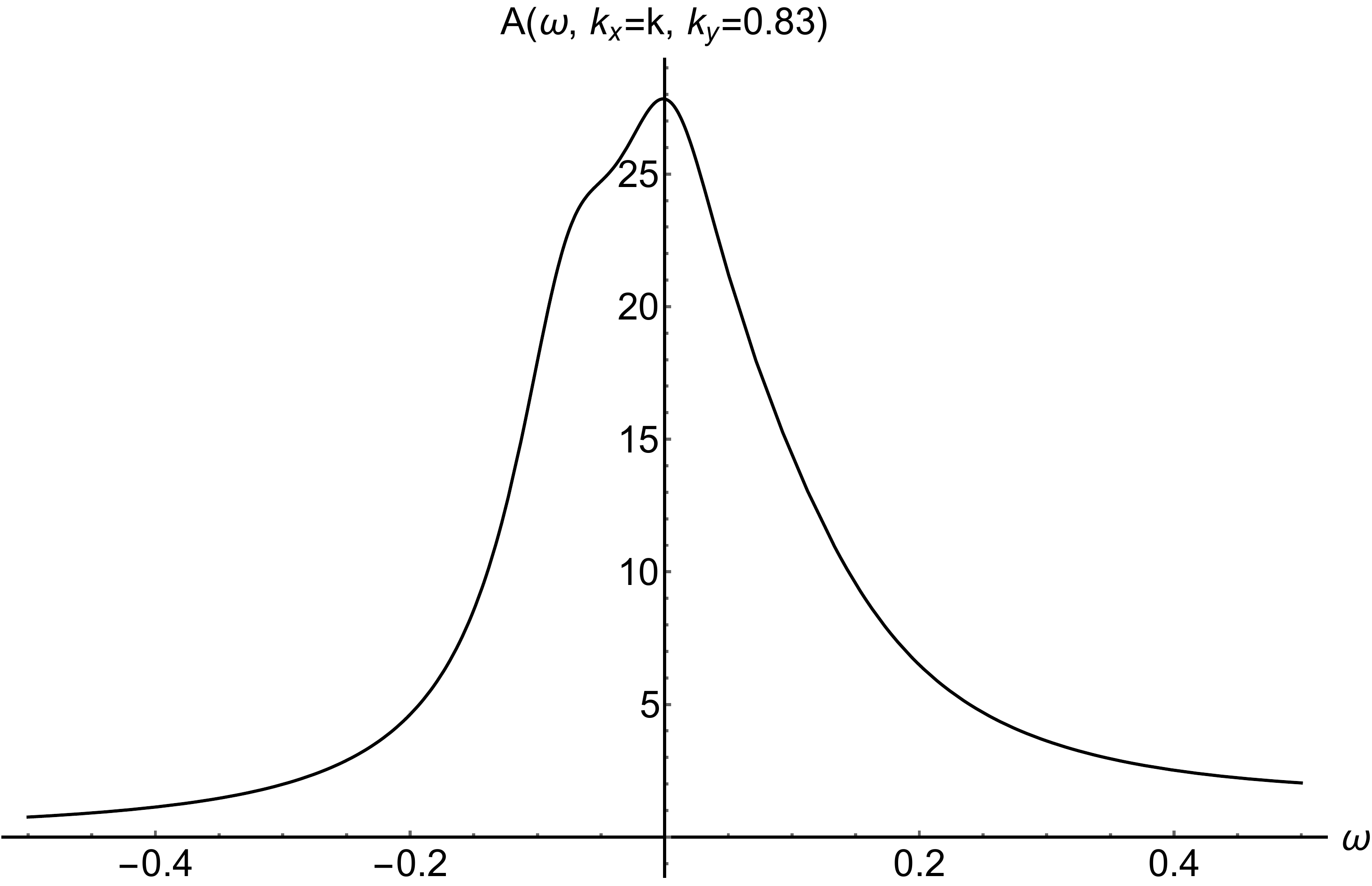} \\
\includegraphics[width=.45\textwidth]{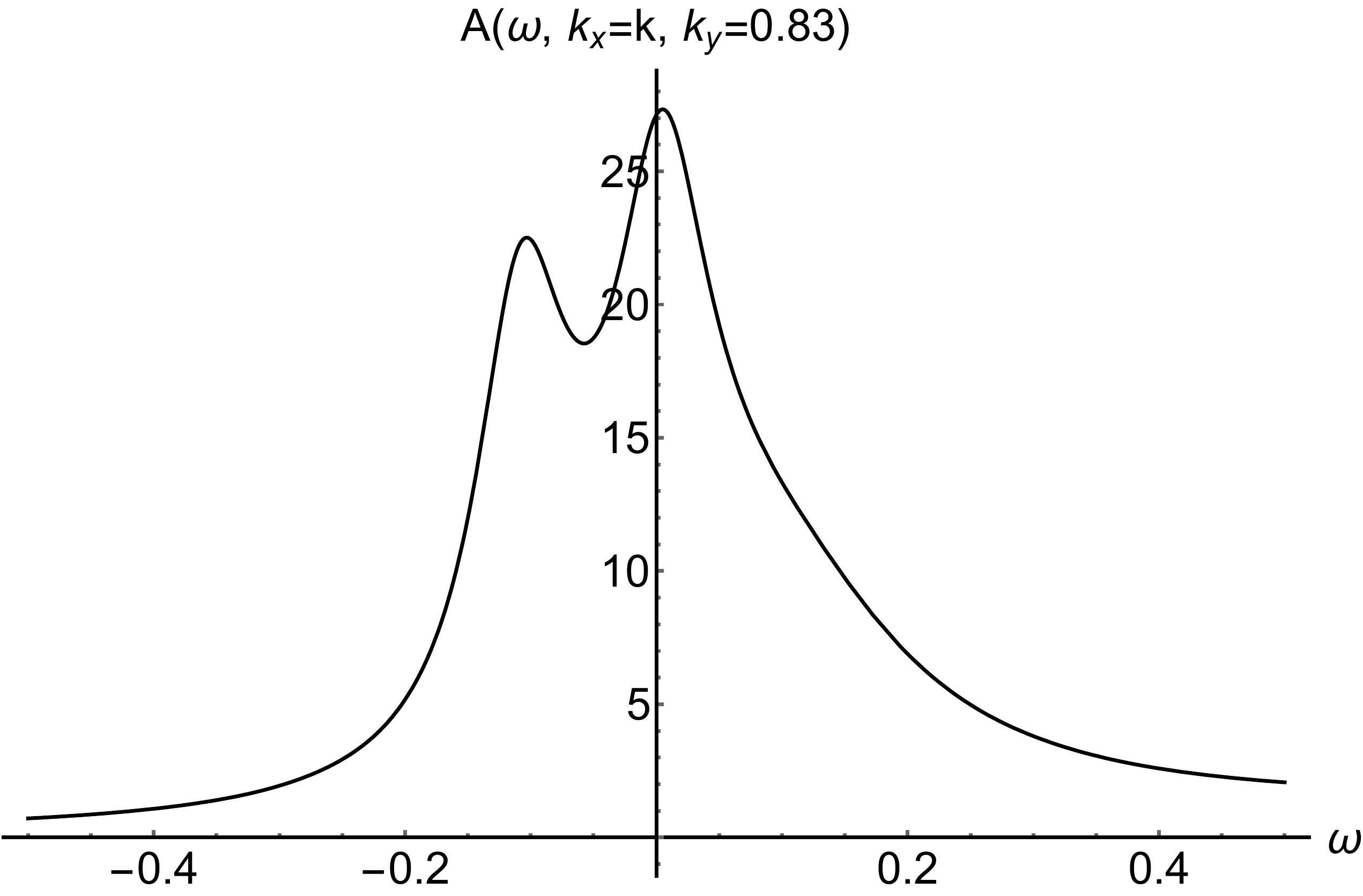}
\caption{Plot of spectral function, $A(\omega, k_x=k, k_y=0.83)$, as a function of $\omega$ for non-Fermi liquid case, $\nu_{k_l}<1/2$, with parameters
$\frac{\eta}{\mu_0^2}=0.25$, $\frac{\eta'}{\mu_0^4}=0.005$, $q=0$,
$\Delta=1$, $q_f=1.15$, $\mu_0=2.2507$, $\frac{T}{\mu_0}=0.0613$ and for $\xi=0.0, 0.25, 0.3$ respectively. } \label{nflgapomega}
\end{figure}

In Fig. \ref{nflgapomega} we plotted the spectral function for
$\nu_{k_l}=0.3134 <1/2$ with the parameters $\frac{\eta}{\mu_0^2}=0.25$,
$\frac{\eta'}{\mu_0^4}=0.005$, $q=0$, $\Delta=1$, $q_f=1.15$,
$\mu_0=2.2507$, $\frac{T}{\mu_0}=0.0613$ for
$\xi=0.0, 0.25, 0.3$. We were unable to observe a gap unless
$\xi\gtrsim 0.25$. This is because the spectral functions
are wide enough to hide the peaks for small values of $\xi$. Also as seen from Fig. \ref{nflgapomega}, the broad peaks correspond to a lack of stable quasiparticles.

\item Finally, for  $\nu_{k_l} = \frac{1}{2}$, the system is in the marginal
Fermi liquid state.  There is still a Fermi surface but the self energy
scaling is not quadratic in $\omega$ as $\Im \Sigma \propto
\omega \ln \omega$. Therefore, near Fermi surface and small $\omega$,
the $\omega$-dependence of the matrix elements of $A$ are $A_{\alpha
l, \alpha l}=\omega+c_2 \omega \ln \omega-v_F(k_l-k_F)+i c_1
\omega \ln \omega$.

\end{itemize}



\section{Conclusion}
\label{DiracConclusion}

We have studied the behavior and properties of a holographic
Fermi liquid system in a spontaneously generated lattice. The Dirac field was calculated in the presence of a finite but small-temperature black hole.
Exploring the holographic fermionic spectral function, we
found a pseudogap at the edge of the Brillouin
zone for the degenerate case, $\nu_{k_l}=\nu_{k_{l'}}$, due to
 interactions between different levels. The magnitude of the
gap increases with increasing order parameter
$\xi$.  However, with large enough $\xi$, the perturbation limit breaks
down.

These results are consistent with the lattice effects on the Fermi
surface due to a periodic potential previously studied in
\cite{LSSZ:2012} and with backreaction  in
\cite{Ling:2013fl}. However, our motivation was unique. Instead
of choosing a modulated scalar potential for the
electromagnetic field at the start, we introduced a higher-derivative
interaction  between a $U(1)$ gauge field and a scalar field.
We used perturbation theory to expand the bulk fields  below
$T_c$, thus obtaining an analytic solution to the coupled system of
Einstein-Maxwell-scalar field equations at first order. We
found a spatially inhomogeneous charge
density spontaneously generated in the boundary theory \cite{Alsup:2013kda}.

We provided analytic and
numerical solutions of the gap behavior in
the presence of a spontaneously generated lattice.  The behavior of the holographic fermionic system was studied by
solving the Dirac equation in the gravitational background of the backreacted
Einstein-Maxwell-scalar system. It will be
interesting to perform a future study of the backreaction of Dirac fields. As it is well known, fermions should not be
treated as elementary fields but as a fluid.  Therefore the
energy-momentum tensor of a perfect fluid must be introduced
into the field equations, leading to an electron star
\cite{Hartnoll:2010gu,Allais:2013lha,Cubrovic:2010bf} with lattice effects.

Another direction of future research is to introduce a dipole
coupling of  fermions to an electromagnetic field
\cite{Edalati:2010ww,Guarrera:2011my,Edalati:2010ge}. There have been some  studies in this direction introducing  a Q-lattice
background \cite{Ling:2014bda,Ling:2015mott}. It will be
interesting to extend these to a fully dynamical
generation of a   Mott gap, and analyze the dynamically generated lattice
effects  as a function of the dipole coupling.

\acknowledgments{ E.\ P.\ is partially supported by
	the Greek Ministry of Education and Religious Affairs, Sport and Culture through the
	ARISTEIA II action of the operational program Education and Lifelong Learning.}

\begin{appendix}

\section{ First-order field modes  }
\label{cmf}

In this appendix, we calculate  the metric function and
Maxwell field modes at first order.
Solving the  Maxwell equations (\ref{maxeq}) at first order, we obtain
\begin{equation}\label{At10eq}
A_{t}^{1,0}(z)=C+\mathbf{a}_t^{1,0}(z) \ , \ \
\mathbf{a}_t^{1,0}(z)=\frac{\mu_0}{4}\int_1^z \frac{dw}{(1-w)^2}\int_1^{w} dw'\ {w'}^{2 \Delta-2} \frac{(1-w')^3}{
	h^2(w')} \mathcal{A}(w')
\end{equation}
where
\begin{equation}
\begin{split}
\mathcal{A}(z)=& \left[\frac{4q^2(1-z)}{h(z)} \left( 1+ \frac{\mu_0^2 z^3(1-z)}{4h(z)} \right)+z\left[\Delta^2+8k^2z^2\eta(1+\Delta)\right] \right] F^2(z) \\&+ z^2 \left[ 2
(\Delta+4 k^2  \eta z^2 ) F(z)+z F'(z) \right] F'(z)
\label{max1}~.
\end{split}
\end{equation}
and the integration constant $C$ remains to be determined.

Having obtained $A_{t}^{1,0}(z)$, we proceed to solve the Einstein
equations to find  $Q_{tt}^{1,0}, Q_{zz}^{1,0}, Q_{xx}^{1,0}, Q_{yy}^{1,0}$.
We find that   $Q_{xz}^{1,0}$ does not appear  in the first-order
equations and can set $Q_{xz}^{1,0}(z)=0$, as a gauge choice. We find the following analytic solutions of  the metric functions
\begin{equation}
\begin{split}
Q_{tt}^{1,0}(z)=&\frac{1}{2}\mathcal{Q}_{1}(1)\left(3- \frac{\mu_0^2}{4}\right)\frac{z^3(1-z)}{h(z)}+\frac{z^3}{ h(z)}\int_1^z dw \left[\frac{12\mathcal{Q}_{1}(w)}{w^4}\right. +
\left. \frac{w^{2\Delta -4} }{4 h^2(w)}\mathcal{Q}_{2}(w)\right]
\\&-\frac{z^4}{ h(z)}\int_1^z dw \left[ \frac{12\mathcal{Q}_{1}(w)}{w^5}\right. +
\left. \frac{w^{2\Delta -5} }{4 h^2(w)}\mathcal{Q}_{2}(w)\right]~, \label{eqQtt10}
\end{split}
\end{equation}
where
\begin{equation}
\mathcal{Q}_{1}(z)=-\frac{1}{2}\int_0^z dw \, w^{2 \Delta +1} \left( \frac{q^2\mu_0^2  (1-w)^2F(w)^2}{h^2(w)}+\frac{ \left(w F'(w)+\Delta  F(w)\right)^2}{w^2}\right)~, \label{mathcalQ1}
\end{equation}
and
\begin{equation}
\begin{split}
\mathcal{Q}_{2}(z)=&  z h^2(z) \left( - z h'(z) +4h( z) \right) (2 \Delta F(z)+ zF'(z))F'(z)\\&+ \left[-h^2(z) \left(2 \left(2 (\Delta -3) \Delta +k^2 z^2 \left( 1+\eta  \mu_0^2 z^4 \right)\right)+\Delta ^2   ( zh'(z) -4 h(z))\right)
\right. \\&
\left. \ \ \ \ \ -\mu_0^2 q^2 z^2 (1-z)^2 (zh'(z)-8h(z)) \right] F^2(z)~. \label{mathcalQ2}
\end{split}
\end{equation}
Similarly, we find
\begin{equation}
Q_{zz}^{1,0}(z)=\frac{1}{2}\mathcal{Q}_{1}(1)\left(3- \frac{\mu_0^2}{4}\right)\frac{z^3(1-z)}{h(z)}+\frac{z^3}{4h(z)}\int_1^z dw \frac{w^{2 \Delta -5} (w-z) }{h^2(w)} \mathcal{Q}_{3}(w)~,\label{eqQzz10}
\end{equation}
where
\begin{equation}
\begin{split}
\mathcal{Q}_{3} (z)= &2 z h \left(3 \Delta  z h h'+2 \Delta  (3 \Delta -4) h^2+2 q^2 \mu_0^2  (1-z)^2 z^2\right) F'F
\\&+\left[h^2 \left(3 \Delta^2 z h'-2 \left(2 (\Delta -3) \Delta +k^2 z^2 \left( 1+\eta  \mu_0^2 z^4 \right)\right)\right)-q^2\mu_0^2 (1-z)^2 z^3 h'\right]F^2
\\&+\left[2 q^2 \mu_0^2  (1-z) z^2 h (2 \Delta (1-z)-1-z)+2 \Delta ^2 (2 \Delta -5) h^3\right]F^2
\\&+4 \Delta  z^2 h^3 F'' \left[\Delta F+z F'\right]
+\left[z^2 h^2 \left(3 z h'+(8 \Delta -6) h\right)\right]{F'}^2~. \label{mathcalQ4}
\end{split}
\end{equation}
The remaining modes are found to be
\begin {equation}
Q_{xx}^{1,0}(z)= -\frac{1}{2} k^2 \int_0^z dw \frac{w^2}{h(w)}\int_1^{w} dw'
 {w'}^{2\Delta-2}( 1-\eta \mu_0^2 {w'}^4 )F^2(w') ~,\label{eqQxx10}
\end{equation}
and, lastly,
\begin{equation}
Q_{yy}^{1,0}(z)=-Q_{xx}^{1,0}(z)~.\label{eqQyy10}
\end{equation}
Having obtained $A_t^{1,0}$ and $Q_{\mu\nu}^{1,0}$, the $Q_{\mu\nu}^{1,1}$ and $A_t^{1,1}$ modes may be deduced
from the remaining  Einstein-Maxwell equations.
These amount to six equations with six unknown
functions. Two of the equations are first order.
The system of equations to be solved numerically is comprised of
\begin{equation}
\begin{split}
&{Q_{xx}^{1,1}}''+{Q_{yy}^{1,1}}''-\frac{ z \left(8 k^2 z-4h'+\mu_0^2 z^3\right)+12 h }{2 z^2
	h} Q_{zz}^{1,1}-\frac{4 k^2}{h} Q_{yy}^{1,1} \\&-\frac{2 k z^2
	h'}{h}Q_{xz}^{1,1} -4 k z^2
{Q_{xz}^{1,1}}'+\frac{2 }{z}{Q_{zz}^{1,1}}'+\frac{ z h'-4h
	}{2 z h} \left( {Q_{xx}^{1,1}} + {Q_{yy}^{1,1}}\right)'-\frac{\mu_0^2 z^2}{2
	h } Q_{tt}^{1,1}\\&-\frac{\mu_0 z^2 (1-z)}{h} {A_{t}^{1,1}}'+\frac{\mu_{0} z^2
	}{h} A_{t}^{1,1}+ z^{2\Delta -2}\mathcal{Q}_4=0~, \label{eq2}
\end{split}
\end{equation}
\begin{equation}
\begin{split}
&(-zh'+4h)\left(Q_{xx}^{1,1}+Q_{yy}^{1,1}\right)'+4 h
{Q_{tt}^{1,1}}'+ \left( -4 h'
+\mu_{0}^2 z^3+\frac{12h}{z} \right) Q_{zz}^{1,1}\\&-4 k z^2(-zh'+4h) Q_{xz}^{1,1}(z)+\left(8 k^2
	z+\mu_0^2 z^3\right)Q_{tt}^{1,1}+8 k^2 z
	Q_{yy}^{1,1}\\&-2 \mu_{0} z^3 A_{t}^{1,1}+2 \mu_{0} (1-z) z^3 {A_{t}^{1,1}}'+ 2z^{2\Delta -1} h\mathcal{Q}_5=0~, \label{eq4}
\end{split}
\end{equation}
\begin{equation}
\begin{split}
{Q_{tt}^{1,1}}'+{Q_{yy}^{1,1}}'+\frac{h'}{2 h} Q_{tt}^{1,1}+\frac{4h- z h'}{2 zh
	}Q_{zz}^{1,1}+\frac{\mu_{0} z^2 (1-z)}{h}A_{t}^{1,1}+z^{2 \Delta -1}\mathcal{Q}_6=0~, \label{eq6}
\end{split}
\end{equation}
\begin{equation}
\begin{split}
&{Q_{tt}^{1,1}}''+{Q_{yy}^{1,1}}''+\frac{ -3 z h'+4h
	}{2 z h}{Q_{tt}^{1,1}}' +\frac{z h'-2h
	}{z h}{Q_{yy}^{1,1}}'-\frac{z h'-4h
	 }{2 z h} {Q_{zz}^{1,1}}'+\frac{\mu_{0}^2 z^2 }{2 h}Q_{tt}^{1,1}\\&+\frac{\mu_{0} z^2 (1-z)}{h}{A_{t}^{1,1}}'-\frac{z \left(
	 8h'+2zh''-\mu_{0}^2 z^3 \right)+12h
	 }{2 z^2 h}Q_{zz}^{1,1}-\frac{\mu_{0} z^2 }{
	h}A_{t}^{1,1}+z^{2 \Delta -2}\mathcal{Q}_7=0~, \label{eq8a}
\end{split}
\end{equation}
\begin{equation}
\begin{split}
&{Q_{tt}^{1,1}}''+{Q_{xx}^{1,1}}''-\frac{ z \left( 8 k^2 z-8h'+2zh''
	-\mu_{0}^2 z^3  \right)+12h
	}{2 z^2 h}Q_{zz}^{1,1}\\&-\frac{8
	k^2-\mu_{0}^2 z^2}{2 h}Q_{tt}^{1,1} -\frac{4 k z^2
	h'}{h}Q_{xz}^{1,1}+\frac{3 z h'-4
	h }{ z h}{Q_{tt}^{1,1}}'+\frac{z h'-2h
	 }{ z h}{Q_{xx}^{1,1}}'-\frac{z h'-4h
	 }{2 z h}{Q_{zz}^{1,1}}'\\&+\frac{\mu_{0} z^2
	(1-z)}{h}{A_{t}^{1,1}}'-\frac{\mu_{0} z^2 }{h}A_{t}^{1,1}-4 k z^2
{Q_{xz}^{1,1}}'+z^{2 \Delta -2}\mathcal{Q}_8=0~, \label{eq10a}
\end{split}
\end{equation}
and
\begin{equation}
\begin{split}
&{A_{t}^{1,1}}''-\frac{4 k^2 }{h}A_{t}^{1,1}-\frac{2
	}{1-z}{A_{t}^{1,1}}'+\frac{2 k \mu_{0} z^2 }{1-z}Q_{xz}^{1,1}\\&+\frac{\mu_{0}
	}{2(1- z)}( {Q_{tt}^{1,1}}- {Q_{xx}^{1,1}}- {Q_{yy}^{1,1}}+
	{Q_{zz}^{1,1}})'-\frac{\mu_{0} z^{2 \Delta -2} \left(q^2-2 \eta  k^2
		z^3 h\right)}{(1-z)h} F^2=0~,
\end{split}
\end{equation}
where
\begin{equation*}
\mathcal{Q}_4= \frac{1}{2} z^{2 } F'^2+\Delta z F
F' + \frac{ \left( (\Delta -3) \Delta-k^2 z^2
	\left( 1+ \eta  \mu_0^2 z^4 \right)
	\right) h+\Delta ^2 h^2+\mu_0^2 q^2 (1-z)^2 z^2}{2
	h^2} F^2~,
\end{equation*}
\begin{equation*}
\mathcal{Q}_5= \frac{1}{2} z^{2 } F'^2+\Delta z F
F' + \frac{ \left( -(\Delta -3) \Delta+k^2 z^2
	\left( 1+ \eta  \mu_0^2 z^4 \right)
	\right) h+\Delta ^2 h^2+\mu_0^2 q^2 (1-z)^2 z^2}{2
	h^2} F^2~,
\end{equation*}
\begin{equation*}
\begin{split}
\mathcal{Q}_6=&\frac{1}{2} F  \left(z F'+\Delta
F\right)~,
\end{split}
\end{equation*}
\begin{equation*}
\mathcal{Q}_7=\frac{1}{2} z^{2 } F'^2+\Delta  z F F'+\frac{   \left((\Delta -3)
	\Delta +k^2 z^2\left(1-\eta  \mu_{0}^2 z^4\right)\right) h+\Delta^2
	h^2-\mu_{0}^2 q^2 (1-z)^2 z^2}{2 h^2}F^2~,
\end{equation*}
\begin{equation*}
\mathcal{Q}_8=\frac{1}{2} z^{2 } F'^2+\Delta   zF F'+\frac{  \left((\Delta -3) \Delta
	-k^2 z^2 \left( 1-\eta  \mu_{0}^2 z^4 \right)\right) h+\Delta ^2
	h^2-\mu_{0}^2 q^2 (1-z)^2 z^2}{2 h^2}F^2~.
\end{equation*}

We can solve \eqref{eq4}
for $Q_{xz}^{1,1}$ and \eqref{eq6} for $Q_{zz}^{11}$. We will also set $Q_{xz}^{1,1}=0$, which is a gauge choice.
 The remaining equations are solved with the boundary conditions
specified in Eqs.\ \eqref{QhorBC}, \eqref{QBbc}, and \eqref{Fbc}.

Notice that some of the modes depend on the integration constant $C$ through the gauge field mode $A_t^{1,0}$. In order to determine $C$, we use the first order
scalar field expanded in Fourier modes (eq.\ \eqref{fourier}).
The scalar field equation for the mode $\phi^{1,0}(z)$
is
\begin{equation}
\begin{split}
{\phi^{1,0}}''&+\frac{z h'-2h }{z
	h}{\phi^{1,0}}'+\frac{- \left( (\Delta -3)\Delta +k^2 z^2\left(1-\eta   \mu_{0}^2
	z^4\right)\right) h+\mu_{0}^2 q^2 (1-z)^2 z^2}{ z^2
	h^2}\phi^{1,0}\\& +\left[ C (\mathcal{C}_0 F) + \mathcal{D}_2 F''+\mathcal{D}_1
F'+\mathcal{D}_0 F \right] z^\Delta  =0~,
\label{firstscalar}
\end{split}
\end{equation}
where the functions
$\mathcal{C}_0$, $\mathcal{D}_2$, $\mathcal{D}_1$,
$\mathcal{D}_0$ are given, respectively, by

\begin{equation}
\begin{split}
\mathcal{C}_0=& \frac{2 \mu_0 \left[\eta  k^2 z^4 h+q^2 (1-z)^2\right]}{ h^2}
~, \label{eqC0}
\end{split}
\end{equation}
\begin{equation}
\mathcal{D}_2=- Q_{zz}^{1,0}-\frac{1}{2} Q_{zz}^{1,1}~,\label{eqD2}
\end{equation}
\begin{equation}
\begin{split}
\mathcal{D}_1=& \frac{ 2(1- \Delta) h - z h'}{z h} Q_{zz}^{1,0} -\frac{ 2 (1+\Delta) h+ z h'}{2 z h} Q_{zz}^{1,1}
\\&+\frac{1}{4} \left( 2 Q_{tt}^{1,0} +Q_{tt}^{1,1}+2 Q_{xx}^{1,0}+Q_{xx}^{1,1}+2 Q_{yy}^{1,0}+Q_{yy}^{1,1}-2 Q_{zz}^{1,0}-Q_{zz}^{1,1} \right)'~,\label{eqD1}
\end{split}
\end{equation}
and
\begin{equation}
\begin{split}
\mathcal{D}_0=& -\frac{ z \left(\eta  k^2
	\mu_0^2 z^5+\Delta h'\right)+\Delta  (\Delta
	 +2)h}{z^2 h} Q_{zz}^{1,0}
+\frac{z \left(k^2 z-\Delta  h'\right)+\Delta   (3-\Delta  )h}{2z^2 h} Q_{zz}^{1,1}
\\&-\frac{ \mu_0^2 \left( \eta  k^2 z^4 h+q^2(1-z)^2\right)}{h^2}  Q_{tt}^{1,0} +\frac{ \left( k^2 h-\mu_0^2 q^2 (1-z)^2 \right)}{2h^2}  Q_{tt}^{1,1} \\&+\frac{2 \mu_0 \left(q^2(1-z)^2+\eta  k^2 z^4 h\right)}{h^2} \mathbf{a}_t^{1,0} +\frac{ \mu_0 (1-z) \left(q^2(1-z)-2 \eta  k^2 z^3 h\right)}{h^2} A_{t}^{1,1} \\&+\frac{ k^2 \left( 1-\eta  \mu_0^2 z^4 \right)}{h}  Q_{xx}^{1,0} +\frac{ k^2 \left( 1- \eta  \mu_0^2 z^4 \right)}{2h} Q_{yy}^{1,1} -\frac{2 \eta  k^2 \mu_0 z^4  (1-z)}{h}{\mathbf{a}_t^{1,0}}'\\&+\frac{\Delta }{4z} \left( 2Q_{tt}^{1,0}-2 Q_{zz}^{1,0}+2Q_{xx}^{1,0}+2Q_{yy}^{1,0}+Q_{tt}^{1,1}-Q_{zz}^{1,1}+Q_{xx}^{1,1}+Q_{yy}^{1,1}\right)'~.
\label{eqD0}
\end{split}
\end{equation}

The integration constant calculated from \eqref{firstscalar}
is
\begin{equation}\label{eqC}
C=-\frac{\int_0^1 dz \, z^{2\Delta} F
	\left[\mathcal{D}_2F''+\mathcal{D}_1 F'+\mathcal{D}_0
	F\right]}{\int_0^1 dz \, z^{2 \Delta} \mathcal{C}_0 F^2}~.
\end{equation}
Our numerical results are displayed in Figs.\ \ref{metricsolutions} and \ref{chargedensity}, and discussed in the surrounding text.

\section{Simplification of the Dirac equation}
\label{spinorrota}
Here we show how a $SO(2)$ rotation of the Dirac field $\psi$ may be conveniently used to extract the effect of the lattice structure on the fermionic spectral function.

The Dirac equation at the critical temperature (eq.\ \eqref{DiracaboveTc}) in terms of spinor components is the system of equations
\bes \sqrt{h} \left( \psi_{+1}^{0, l} \right)' + \frac{m_f}{z} \psi_{+1}^{0, l} + \left[ k_x+2k l - \frac{\mu_0 q_f (1-z)+\omega}{\sqrt{h}} \right] \psi_{+2}^{0, l} - k_y \psi_{-2}^{0, l} &=& 0~, \nonumber\\
 \sqrt{h} \left( \psi_{+2}^{0, l}\right)' - \frac{m_f}{z} \psi_{+2}^{0, l} + \left[ k_x+2k l + \frac{\mu_0 q_f (1-z)+\omega}{\sqrt{h}} \right] \psi_{+1}^{0, l} - k_y \psi_{-1}^{0, l} &=& 0~, \nonumber\\
 \sqrt{h} \left( \psi_{-1}^{0, l}\right)' + \frac{m}{z} \psi_{-1}^{0, l} + \left[ -k_x -2k l- \frac{\mu_0 q_f (1-z)+\omega}{\sqrt{h}} \right] \psi_{-2}^{0, l} - k_y \psi_{+2}^{0, l} &=& 0~, \nonumber\\
 \sqrt{h} \left( \psi_{-2}^{0, l}\right)' - \frac{m_f}{z} \psi_{-2}^{0, l} + \left[ -k_x - 2k l + \frac{\mu_0 q_f (1-z)+\omega}{\sqrt{h}} \right] \psi_{-1}^{0, l} - k_y \psi_{+1}^{0, l} &=& 0~. \label{eq1}  \ees
Combining the first and third equations into
\bea\label{eq22} && \sqrt{h} (\psi_{+1}^{0, l} - \lambda \psi_{-1}^{0, l} )' +
\frac{m_f}{z} (\psi_{+1}^{0, l}
 - \lambda \psi_{-1}^{0, l})- \left[  \frac{\mu_0 q_f (1-z)+\omega}{\sqrt{h}} \right] (\psi_{+2}^{0, l} -\lambda \psi_{-2}^{0, l} )\nonumber \\
 && +(k_x +\lambda k_y) \psi_{+2}^{0, l} + (-k_y +\lambda k_x) \psi_{-2}^{0, l} =
 0~,
 \eea
with the choice $\lambda =  \tan\frac{\theta}{2}$, where $\tan\theta = \frac{k_y}{k_x+2kl}$, we obtain
\be \sqrt{h}
\left( \tilde{\psi}_{+1}^{0, l}\right)' +
\frac{m_f}{z} \tilde{\psi}_{+1}^{0, l} +
\left[  k_l - \frac{\mu_0 q_f (1-z)+\omega}{\sqrt{h}} \right]
\tilde{\psi}_{+2}^{0, l}  = 0~,
\label{eq44}\ee
which is identical to the first equation in \eqref{eq1} with $k_x+2kl$ replaced by $k_l = \sqrt{(k_x+2kl)^2+k_y^2}$ and $k_y
=0$, and we defined
\be\label{eqB4} \tilde{\psi}^{0, l}_{+s}=\cos\frac{\theta}{2}\psi^{0, l}_{+s}-\sin\frac{\theta}{2} \psi^{0, l}_{-s}~.
\ee Similarly, we obtain the linear combination of the second and fourth equations in the system of equations \eqref{eq1}, \be
 \sqrt{h} \left( \tilde{\psi}_{+2}^{0, l}\right)' - \frac{m_f}{z} \tilde{\psi}_{+2}^{0, l} + \left[  k_l +
 \frac{\mu_0 q_f (1-z)+\omega}{\sqrt{h}} \right] \tilde{\psi}_{+1}^{0, l} = 0~. \label{eq5}  \ee
 For the linearly independent combination
 \be\label{eqB6} \tilde{\psi}^{0, l}_{-s}=\cos\frac{\theta}{2} \psi^{0, l}_{-s}+ \sin\frac{\theta}{2} \psi^{0, l}_{+s}~,
 \ee
 we obtain two more equations,
 \bea \sqrt{h} \left( \tilde{\psi}_{-1}^{0, l}\right)' +
\frac{m_f}{z} \tilde{\psi}_{-1}^{0, l} + \left[ - k_l -
\frac{\mu_0 q_f (1-z)+\omega}{\sqrt{h}} \right] \tilde{\psi}_{-2}^{0, l}  &=& 0~,\nonumber \\ \sqrt{h} \left( \tilde{\psi}_{-2}^{0, l}\right)' - \frac{m_f}{z} \tilde{\psi}_{-2}^{0, l} + \left[ -
k_l + \frac{\mu_0 q_f (1-z)+\omega}{\sqrt{h}} \right] \tilde{\psi}_{-1}^{0, l}
  &=& 0~. \label{eq77} \eea
  Thus with the $SO(2)$ rotation of angle $\frac{\theta}{2}$ (eqs.\ \eqref{eqB4} and \eqref{eqB6}), we obtain a simplified system of equations (eqs.\ \eqref{eq44}, \eqref{eq5}, and \eqref{eq77}) in which the modes $\tilde{\psi}_+$ and $\tilde{\psi}_-$ are decoupled.

\end{appendix}


\begin{thebibliography}{99}

\bibitem{Abrahams}
C. M. Varma, P. B. Littlewood, and S. Schmitt-Rink, E. Abrahams
and A. E. Ruckenstein,  "Phenomenology of the Normal State of Cu-O
High-Temperature Superconductors", Phys. Rev. Lett. 63, 1996,
(1989).

\bibitem{Faulkner:2009am}
  T.~Faulkner, G.~T.~Horowitz, J.~McGreevy, M.~M.~Roberts and D.~Vegh,
  ``Photoemission 'experiments' on holographic superconductors,''
  JHEP {\bf 1003}, 121 (2010)
  [arXiv:0911.3402 [hep-th]].

\bibitem{Maldacena:1997re}
J.~M.~Maldacena, \emph{The large N limit of superconformal field
theories and supergravity}, Adv.\ Theor.\ Math.\ Phys.\  {\bf 2}
(1998) 231 [Int.\ J.\ Theor.\ Phys.\  {\bf 38} (1999) 1113].

\bibitem{Gubser:2002}
 S. S. Gubser, I. R. Klebanov and A. M.
Polyakov, \emph{A semiclassical limit of the gauge string
correspondence}, Nucl. Phys. B\textbf{ 636} (2002) 99.

\bibitem{Witten:1998} E. Witten, \emph{Anti-de Sitter space and
holography}, Adv. Theor. Math. Phys. \textbf{2} (1998) 253.

\bibitem{Gubser:2009dt}
  S.~S.~Gubser, F.~D.~Rocha and P.~Talavera,
  ``Normalizable fermion modes in a holographic superconductor,''
  JHEP {\bf 1010}, 087 (2010)
  [arXiv:0911.3632 [hep-th]].

\bibitem{Nitti:2014fsa}
  F.~Nitti, G.~Policastro and T.~Vanel,
  ``Polarized solutions and Fermi surfaces in holographic Bose-Fermi systems,''
  JHEP {\bf 1412}, 027 (2014)
  [arXiv:1407.0410 [hep-th]].

\bibitem{LSSZ:2012}
  Y.~Liu, K.~Schalm, Y.~W.~Sun, and J.~Zaanen,
  ``Lattice potentials and fermions in holographic non Fermi-liquids: hybridizing local quantum criticality,''
 JHEP {\bf 1210}, 036 (2012)
  [arXiv:1205.5227v2 [hep-th]].

\bibitem{Ling:2013fl}
  Y.~Ling, C.~Niu and J.~-P.~Wu and Z.~-Y.~Xian and H.~Zhang,
  ``Holographic fermionic liquid with lattices,''
 JHEP {\bf 1307}, 045 (2013)
  [arXiv:1304.2128 [hep-th]].

  \bibitem{Maeda:2011pk}
  K.~Maeda, T.~Okamura and J.~-i.~Koga,
  ``Inhomogeneous charged black hole solutions in asymptotically anti-de Sitter spacetime,''
  Phys.\ Rev.\ D {\bf 85}, 066003 (2012)
  [arXiv:1107.3677 [gr-qc]].

\bibitem{Aperis:2010cd}
  A.~Aperis, P.~Kotetes, E.~Papantonopoulos, G.~Siopsis, P.~Skamagoulis and G.~Varelogiannis,
  ``Holographic Charge Density Waves,''
  Phys.\ Lett.\ B {\bf 702}, 181 (2011)
  [arXiv:1009.6179 [hep-th]].

\bibitem{Flauger:2010tv}
  R.~Flauger, E.~Pajer and S.~Papanikolaou,
  ``A Striped Holographic Superconductor,''
  Phys.\ Rev.\ D {\bf 83}, 064009 (2011)
  [arXiv:1010.1775 [hep-th]].


\bibitem{Hutasoit:2012ib}
  J.~A.~Hutasoit, G.~Siopsis and J.~Therrien,
  ``Conductivity of Strongly Coupled Striped Superconductor,''
  arXiv:1208.2964 [hep-th].

\bibitem{Erdmenger:2013zaa}
  J.~Erdmenger, X.~H.~Ge and D.~W.~Pang,
  ``Striped phases in the holographic insulator/superconductor transition,''
  JHEP {\bf 1311}, 027 (2013)
  [arXiv:1307.4609 [hep-th]].

\bibitem{Donos:2013eha}
  A.~Donos and J.~P.~Gauntlett,
  ``Holographic Q-lattices,''
  JHEP {\bf 1404}, 040 (2014)
  [arXiv:1311.3292 [hep-th]].

\bibitem{Horowitz:2012ky}
  G.~T.~Horowitz, J.~E.~Santos and D.~Tong,
  ``Optical Conductivity with Holographic Lattices,''
  JHEP {\bf 1207}, 168 (2012)
  [arXiv:1204.0519 [hep-th]].

\bibitem{Horowitz:2012gs}
  G.~T.~Horowitz, J.~E.~Santos and D.~Tong,
  ``Further Evidence for Lattice-Induced Scaling,''
  JHEP {\bf 1211}, 102 (2012)
  [arXiv:1209.1098 [hep-th]].

\bibitem{Alsup:2013kda}
  J.~Alsup, E.~Papantonopoulos, G.~Siopsis and K.~Yeter,
  ``Spontaneously Generated Inhomogeneous Phases via Holography,''
  Phys.\ Rev.\ D {\bf 88}, no. 10, 105028 (2013)
  [arXiv:1305.2507 [hep-th]].

\bibitem{Faulkner:2010da}
  T.~Faulkner, N.~Iqbal, H.~Liu, J.~McGreevy and D.~Vegh,
  ``From Black Holes to Strange Metals,''
  arXiv:1003.1728 [hep-th].



  \bibitem{BHY2010}
  F.~Benini, C.~P.~Herzog and A.~Yarom,
  ``Holographic Fermi arcs and a d-wave gap,''
  Phys.\ Lett.\ B {\bf 701}, 626 (2011)
  [arXiv:1006.0731 [hep-th]].



\bibitem{Faulkner:2009wj}
  T.~Faulkner, H.~Liu, J.~McGreevy and D.~Vegh,
  ``Emergent quantum criticality, Fermi surfaces, and AdS(2),''
  Phys.\ Rev.\ D {\bf 83}, 125002 (2011)
  [arXiv:0907.2694 [hep-th]].

\bibitem{Faulkner:2011tm}
  T.~Faulkner, N.~Iqbal, H.~Liu, J.~McGreevy and D.~Vegh,
  ``Holographic non-Fermi liquid fixed points,''
  Phil.\  Trans.\  Roy.\  Soc.\ A {\bf  369}, 1640 (2011)
  [arXiv:1101.0597 [hep-th]].


\bibitem{Hartnoll:2010gu}
  S.~A.~Hartnoll and A.~Tavanfar,
  ``Electron stars for holographic metallic criticality,''
  Phys.\ Rev.\ D {\bf 83}, 046003 (2011)
  [arXiv:1008.2828 [hep-th]].

\bibitem{Allais:2013lha}
  A.~Allais and J.~McGreevy,
  ``How to construct a gravitating quantum electron star,''
  Phys.\ Rev.\ D {\bf 88}, no. 6, 066006 (2013)
  [arXiv:1306.6075 [hep-th]].

\bibitem{Cubrovic:2010bf}
  M.~Cubrovic, J.~Zaanen and K.~Schalm,
  ``Constructing the AdS Dual of a Fermi Liquid: AdS Black Holes with Dirac Hair,''
  JHEP {\bf 1110}, 017 (2011)
  [arXiv:1012.5681 [hep-th]].

\bibitem{Edalati:2010ww}
  M.~Edalati, R.~G.~Leigh and P.~W.~Phillips,
  ``Dynamically Generated Mott Gap from Holography,''
  Phys.\ Rev.\ Lett.\  {\bf 106}, 091602 (2011)
  [arXiv:1010.3238 [hep-th]].

\bibitem{Guarrera:2011my}
  D.~Guarrera and J.~McGreevy,
  ``Holographic Fermi surfaces and bulk dipole couplings,''
  [arXiv:1102.3908 [hep-th]].

\bibitem{Edalati:2010ge}
  M.~Edalati, R.~G.~Leigh, K.~W.~Lo and P.~W.~Phillips,
  ``Dynamical Gap and Cuprate-like Physics from Holography,''
  Phys.\ Rev.\ D {\bf 83}, 046012 (2011)
  [arXiv:1012.3751 [hep-th]].



\bibitem{Ling:2014bda}
  Y.~Ling, P.~Liu, C.~Niu, J.~P.~Wu and Z.~Y.~Xian,
  ``Holographic fermionic system with dipole coupling on Q-lattice,''
  JHEP {\bf 1412}, 149 (2014)
  [arXiv:1410.7323 [hep-th]].

\bibitem{Ling:2015mott}
  Y.~Ling, P.~Liu, C.~Niu, J.~P.~Wu,
  ``Building a doped Mott system by holography,''
  [arXiv:1507.02514 [hep-th]].



\end{thebibliography}
\end{document}